\pdfoutput=1
\documentclass[compsoc,12pt,journal,onecolumn,letterpaper]{IEEEtran}

\usepackage{setspace}
\IEEEoverridecommandlockouts
\usepackage{amsmath,amsfonts}
\usepackage{array}
\usepackage{bm}

\usepackage[noend]{algpseudocode}

\usepackage{algorithmicx,algorithm}

\usepackage{textcomp}
\usepackage{stfloats}
\usepackage{url}
\usepackage{verbatim}
\usepackage{graphicx}
\usepackage{cite}

\usepackage{graphics,amsmath,amssymb}
\usepackage{subfigure}

\newtheorem{theorem}{Theorem}[section]

\newtheorem{definition}[theorem]{Definition}

\newcommand{\rmnum}[1]{\romannumeral #1}
\newcommand{\Rmnum}[1]{\uppercase\expandafter{\romannumeral #1\relax}}

\DeclareMathSymbol{\mlq}{\mathord}{operators}{``}
\DeclareMathSymbol{\mrq}{\mathord}{operators}{`'}
\usepackage{tablefootnote}

\usepackage{mathrsfs}
\usepackage{diagbox}
\usepackage{amssymb}
\usepackage{color, soul}
\usepackage{tikz}
\usepackage{tabularx}

\usepackage[commandnameprefix=always]{changes}

\usepackage{threeparttable}
\usepackage{booktabs}

\usepackage{enumitem}

\usepackage{caption}

\usepackage{color}

\usepackage{changes}

\begin{document}

\title{A Near-Optimal Category Information Sampling in RFID Systems}

\author{
Xiujun Wang,
Zhi Liu,~\IEEEmembership{Senior Member,~IEEE,}
Xiaokang Zhou,
Yong Liao,
Han Hu,
Xiao Zheng,
Jie Li,~\IEEEmembership{Senior Member,~IEEE}
\IEEEcompsocitemizethanks{
\IEEEcompsocthanksitem

Xiujun Wang is with the School of Computer Science, Anhui University of Technology, Ma'anshan 243032, China, and also with the
Anhui Engineering Research Center for Intelligent Applications and Security of Industrial Internet, Ma'anshan 243032, China (e-mail: wxj@mail.ustc.edu.cn).
}
}

\IEEEtitleabstractindextext{
\begin{abstract}
In many RFID-enabled applications, objects are classified into different categories, and the information associated with each object's category (called category information) is written into the attached tag, allowing the reader to access it later.
The category information sampling in such RFID systems, which is to randomly choose (sample) a few tags from each category and collect their category information, is fundamental for providing real-time monitoring and analysis in RFID.
However, to the best of our knowledge, two technical challenges, i.e., how to guarantee a minimized execution time and
\textcolor{black}{reduce collection failure caused by missing tags, remain unsolved for this problem.}
In this paper, we address these two limitations by considering how to use the shortest possible time to sample a different number of random tags from each category and collect their category information sequentially in small batches.
In particular,
we first obtain a lower bound on the execution time of any protocol that can solve this problem.
Then, we present a near-\underline{OPT}imal \underline{C}ategory information sampling protocol
(OPT-C) that solves the problem with an execution time close to the lower bound.
Finally, extensive simulation results demonstrate the superiority of OPT-C over existing protocols, while real-world experiments validate the practicality of OPT-C.
\end{abstract}

}

\maketitle
\IEEEdisplaynotcompsoctitleabstractindextext
\IEEEpeerreviewmaketitle

\section{Introduction}
Radio frequency identification (RFID) technology,
with many shining advantages such as non-line-of-sight detection and low manufacturing costs,
has been used in many applications,
ranging from object tracking and positioning \cite{9239290,ngamakeur2020survey}
to supply chain control and scheduling \cite{liu2022rf,wang2021artifact,cai2022tags}.
Typically,
these applications track objects belonging to various categories, such as types of drugs in a pharmacy, topics of books in a library, or brands of food in a supermarket.
Whenever a tag is attached to one of these objects, information related to the category of the object (called category information) can be written to this tag to make it readily available to RFID readers in an offline manner (without consulting a remote reference database) \cite{liu2018efficient,chu2021efficient}.
Such category information on tags can be \textcolor{black}{either} static data (e.g., the brand of a garment)
or dynamic data (e.g.,
\textcolor{black}{continuous sensory readings from a tag's integrated sensor)}
\cite{WISPTag, xie2017minimal,9086079}.

{\color{black}{
In RFID systems for such applications, tags are systematically classified, with all tags within the same category sharing identical category information.
In an ideal scenario, assuming all tags are present and functioning correctly in an interference-free environment, querying a single random tag per category would provide a comprehensive view of the category information. However, practical realities necessitate a departure from this ideal. Overreliance on a single random tag per category is often insufficient, primarily due to the potential absence of the selected tag. Missing tags may result from various factors, such as their location beyond the reader's detection range, signal interference, or hardware malfunctions \cite{chu2021efficient,liu2022revisiting}. For instance, a tag may become unresponsive when obstructed by signal-shielding materials like metal surfaces or when its backscattered signal lacks sufficient power or becomes corrupted, rendering it undetectable \cite{ahson2017rfid,alcaraz2012rfid}.

Therefore, in this paper, we consider the practical RFID system with missing tags
and study the Category Information Sampling Problem \textcolor{black}{(the CIS problem)}, which is to use the
shortest possible time to randomly choose a \textcolor{black}{user-specific number of}
tags from each category and read them.
Here, we afford users
the flexibility to specify
in advance how many tags they would like to
randomly choose from each category,
which can be used to mitigate collection failures caused by missing tags
or for other practical objectives (e.g., data consistency verification and the estimation of category-specific statistics).
This problem holds significant relevance to real-time monitoring of tagged objects and the on-the-fly analysis of their category data within RFID systems.
For instance, consider a supermarket that organizes food products based on their expiration dates, with these dates recorded on attached tags as category information. Due to signal interference or hardware issues, certain tags might be missing. In such cases, swift sampling of expiration dates from randomly selected tags within each category enables real-time tracking of food quality.
Similarly, in a bustling logistics center, shipping items are classified by their destination addresses, written on attached tags as category information. Given the rapid movement of these items, some tags may be missed if they exit the reader's detection area. To ensure timely collection and analysis of shipping addresses, it's imperative to sample multiple tag addresses from each category, rather than relying on just one tag.
Numerous real-world situations present similar contexts where multiple tags convey identical or similar data, necessitating immediate tracking and analysis
\cite{liu2018efficient,chu2021efficient}.
In addition, it is essential to note that various practical RFID scenarios may prompt readers to read multiple tags randomly selected from each category, going beyond the core objective of addressing missing tags. These supplementary purposes encompass tasks such as verifying data consistency and estimating category-specific statistics
\cite{chu2021efficient,xie2017minimal,xiao2019protocol}.

In essence, the investigation of the
\textcolor{black}{CIS problem}
within RFID systems holds the potential to significantly enhance the efficiency and reliability of RFID systems, catering to a wide array of critical applications. This empowers users with the capability to fine-tune their tag selection strategies to address challenges like missing tags and extend their applications to encompass broader objectives such as data validation and statistical estimation.
}
}

\subsection{Prior Art and Limitations}

Numerous protocols exist for extracting data from a tag subset, denoted as $P$, in RFID systems \cite{xie2017minimal,9086079,qiao2015tag,liao2019abnormal,zhao2018information,wang2021efficient}. These protocols often employ Bloom filters or their variants to distinguish the desired subset $P$ from the entire tag population and subsequently retrieve information from the tags within $P$.
For example, ETOP \cite{qiao2015tag} employs multiple Bloom filters to filter out non-target tags within the subset $P$.
Similarly, TIC \cite{9086079} reads a subset $P$ of tags in multi-reader RFID systems, using a single-hash Bloom filter to eliminate irrelevant tags.
\textcolor{black}{However, it is important to note that transmitting Bloom filters and addressing hash collisions can still be time-consuming \cite{lumetta2007using}.}


To the best of our knowledge, for RFID systems with multiple categories of tags,
TPS \cite{liu2018efficient} and ACS \cite{chu2021efficient}
are the only two works that collect category information from randomly sampled tags.
TPS \cite{liu2018efficient}
\textcolor{black}{capitalizes on}
the fact that all tags of
the same category have \textcolor{black}{a common}
category-ID
(indicating which category they belong to)
to collect category information from one random tag of each category.
Specifically, it hashes all tags into slots of a
communication frame based on their category-ID and
finds those homogeneous slots that are occupied only by tags of the same category.
Since the category information of the tags
in the homogeneous slots is the same,
it selects and reads one tag in each homogeneous slot.
However, TPS takes a long time to execute because it transmits many useless empty slots in each frame.
ACS \cite{chu2021efficient} follows TPS, but uses arithmetic coding to compress
each communication frame
to reduce transmission costs.
With ACS, the reader compresses a frame into a binary fraction and
sends it to all tags;
each tag decompresses the fraction to recover the frame.
Even so, the execution time of ACS is still long and unsatisfactory since
it compresses many useless empty slots in each frame;
and the arithmetic encoding is too complex for ordinary RFID tags to support \cite{lin2013enhanced}.


{\color{black}{Ultimately, existing tag sampling protocols suffer from three primary drawbacks:

{\textbf{1. Absence of Execution Time Benchmark:}
There is no known lower bound on the
execution time of any protocol that can solve the
\textcolor{black}{CIS problem.}
Establishing such lower bound is crucial for measuring protocol performance and determining the potential for improvement.}

{\textbf{2. High Execution Times:}
Current sampling protocols exhibit unacceptably high execution times, failing to meet the lower bound. This is particularly problematic for RFID systems, where efficient category information retrieval from
numerous tags is vital for real-time monitoring and analysis.
Achieving execution times close to the lower bound enables us to predict the minimum time required for specific sampling tasks and allocate precise time windows to optimize system efficiency.}

{\textbf{3. Neglect of Missing Tags:}
Existing protocols primarily focus on randomly selecting one tag from each category. However, this approach often falls short because some tags in each category may be missing due to factors like moving out of the reader's range, hardware issues, or signal interference. Consequently, it is more practical to sample a user-specified number of tags from each category, a reasonable requirement that the current protocol inadequately address.
}}}

{\color{black}{In addition,
it's worth noting that
several concurrent tag identification protocols
\cite{wang2012efficient, hu2015laissez, jin2019fliptracer,benedetti2022robust}
have been developed to recover signals
from collided tags using
parallel decoding techniques and specialized equipment like USRP.
For example, the Buzz protocol \cite{wang2012efficient} is a pioneering approach that
effectively coordinates tags to measure their channel coefficients
and reconstructs the transmitted signal of each tag.
In recent \cite{benedetti2022robust}, a novel protocol named CIRF is
proposed to address challenges posed by noisy channels,
which can introduce errors when recovering tag signals.
CIRF leverages multiple antennas to enhance robustness
against noise and uses block sparsity-based optimization techniques
to restore signals from collided tags.
However, these advanced
protocols excel in identifying all tags at the physical
layer and primarily focus on comprehensive tag identification
rather than random sampling from a subset of tags.\footnote{\textcolor{black}{We recognize the extensive research in classical statistics on optimizing device
subset sampling for failure identification or estimation purposes
\cite{mathews2010sample,lenth2001some,kleyner1997bayesian,de2007using}.
These studies often apply Bayesian methods to assess statistical similarities among various devices,
with the aim of determining optimal sample sizes for tasks
such as failure detection and estimation.
However, it's crucial to emphasize the distinction between our
\textcolor{black}{CIS problem}
and these classical approaches.
In our scenario, we do not engage in the optimization of sample size
$c_k$ for identifying missing tags within each category $P_k$
and do not assume any prior knowledge of statistical tag similarities.
Instead, our objective is to generate a simple random sample (SRS) of size $c_k$
for each category $P_k$, where $c_k$ is predetermined by system users,
and to assign unique reporting orders to the sampled tags.}}}}

\subsection{\textcolor{black}{Technical Challenges and Contributions}}\label{sec_proposed_solution}

{\color{black}{This paper addresses the Category Information Sampling Problem
within the context of RFID systems, where the tag population
is categorized into $K$ distinct subsets denoted as
$P_1, P_2,..., P_K$.
The primary objective is to randomly
select a user-defined number of tags, denoted as $c_k$,
from each category $P_k$ and assign them unique reporting
orders for sequential reading while minimizing the execution time.
The parameter $c_k$, called the reliability number,
serves as a user-adjustable parameter aimed at mitigating
collection failures caused by missing tags.
Additionally, it's worth mentioning that the user-adjustable parameter, $c_k$, holds a broader spectrum of applications beyond addressing missing tag issues, as it could be employed for tasks such as category statistics estimation and security enhancement \cite{chu2021efficient, xie2017minimal, 9086079, liu2022revisiting, xiao2019protocol}.}}

\begin{figure}[!t]
\centering
\includegraphics[height=4.5cm]{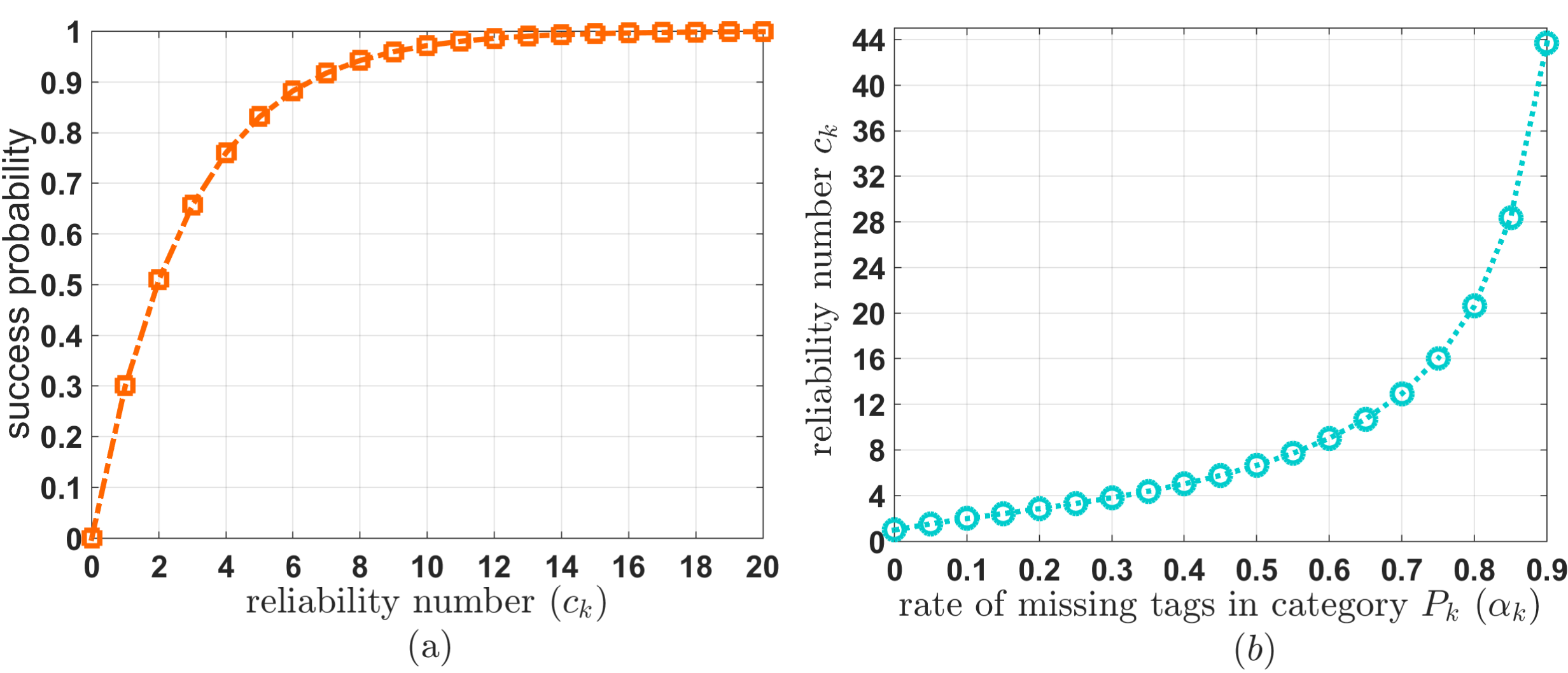}
\caption{Ensuring High Success Probability with a Small Reliability Number.
\footnotesize{Note: (a) The success probability demonstrates rapid growth as the reliability number $c_k$ increases, even when a substantial portion ($70\%$)
of tags in $P_k$ are missing ($\alpha_k=0.7$).
(b)  This graph illustrates the reliability number $c_k$ that can secure a high success probability of 0.99, as $\alpha_k$
varies from $0.05$ to $0.9$.}}
\label{fig-reqc}
\end{figure}

{\color{black}{
Please note that the significance of selecting a suitable value for $c_k$ becomes apparent when considering the issue of missing tags. Sampling a single tag from category $P_k$ carries the risk of encountering missing tags. Hence, users can set $c_k$ to substantially increase the success probability, which represents the likelihood that at least one of the $c_k$ randomly selected tags is not missing, as illustrated in Fig. \ref{fig-reqc}.\footnote{\textcolor{black}{While our paper primarily focuses on efficiently selecting and assigning unique reporting orders to $c_k$ random tags from each category $P_k$ for sequential reading, it's vital to clarify how users can choose a smaller value for $c_k$ to ensure a high success probability when addressing missing tag issues.
First and foremost, it's noteworthy that the rate of missing tags, denoted as $\alpha_k$ (representing the percentage of missing tags within $P_k$), typically remains below $0.9$.
This observation is substantiated by continuous monitoring conducted by RFID readers within their operational range, which serves as a critical measure to mitigate the risks of substantial economic losses and security threats \cite{liu2022revisiting, zhu2020collisions, 9082674, 8948356}.
As $c_k$ increases, the success probability ($1 - \alpha_k^{c_k}$) experiences rapid escalation. Even in an extreme scenario with
$\alpha_k = 0.9$, selecting $c_k = 44$ guarantees a high success probability of $0.99$.
Second, $\alpha_k$ can be efficiently estimated through an RFID counting protocol with a minor time cost $O(\log(\log(|P_k|)))$ \cite{zhou2014understanding}.
These two factors enable users to practically opt for a smaller $c_k$ to maintain a high success probability.
We provide an analytical solution for determining the value of $c_k$ corresponding to any desired success probability in Appendix A.
}}

Addressing the studied problem entails overcoming two major challenges to achieve a near-optimal solution:

\textbf{Challenge 1: Deriving a Non-Trivial Lower Bound on Execution Time}

Previous research on
\textcolor{black}{the CIS problem}
has proposed
protocols for addressing it \cite{liu2018efficient,chu2021efficient}.
However, these works have not ventured into the domain of lower bound
derivation, which serves as a fundamental benchmark for assessing the performance
of these protocols. Deriving this lower bound necessitates addressing two technical problems:\\
\emph{\textbf{(p-1) Analyzing Required Information for
Random Tag Selection and Reporting Orders Assignment:}
This analysis entails dissecting the information that the reader
must transmit to achieve the random selection of a user-defined
number of tags from each category and assign unique reporting orders
to each selected tag.
It aims to identify the essential data within
the reader's broadcasted message and how it contributes to solving the
\textcolor{black}{CIS problem}
at the tag side.}\\
\emph{\textbf{(p-2) Determining Minimum Bits for Information Representation:}
This step involves ascertaining the minimum number of bits to
represent the information needed for (p-1),
considering all possible tag populations and categories.
It seeks to maximize bit utilization while accommodating
constraints arising from the problem's inherent nature and
practical considerations within RFID systems.}

It's worth noting that previous research in this field has not provided solutions to these intricate challenges. Given the practical significance of
\textcolor{black}{the CIS problem,}
we are motivated to tackle these problems. Unlike conventional RFID research, which relies on established lower bounds from communication complexity theory \cite{liu2022revisiting,zhou2014understanding,rao2020communication}, our work is uniquely demanding, as we must develop lower bounds from scratch due to the absence of pre-existing references.

\textbf{Challenge 2: Achieving Near-Optimal Execution Time
in Solving the Problem}

Prior studies \cite{liu2018efficient,chu2021efficient} have proposed protocols for randomly selecting a single
tag from category $P_k$. However, these protocols adopt the
global hashing approach,
which slot all tags in the entire population
into different communication slots but
exploit only the homogeneous slots (a small fraction of all slots) to select tags and read their category information.
This method inevitably leads to a high number of collisions and empty slots due to hash collisions, resulting in significant
transmission overhead. Consequently, these protocols fall short of approaching the derived lower bound on execution time.


In contrast, our proposed two-stage protocol, OPT-C, departs from this global hashing approach. OPT-C comprises OPT-C1 and OPT-C2. OPT-C1 rapidly selects slightly more than $c_k$ tags from each category $P_k$, while OPT-C2 refines the selection process to choose exactly $c_k$ tags and assign unique reporting orders.
These stages introduce two technical problems:\\
\emph{\textbf{(p-3) Designing a Bernoulli Trial for OPT-C1:} OPT-C1 necessitates the design of a Bernoulli trial capable of
swiftly selecting slightly more than $c_k$ tags randomly from category $P_k$. Each tag simultaneously receives a threshold value $\bm{\tau}$ from the reader and uses its local hash value to determine its selection status. This process entails performing a Bernoulli trial in parallel on all tags within each category.
However, the number of tags selected by a Bernoulli trial is a random variable, not a fixed number. Consequently, this aspect
poses a challenge, demanding the careful design of a Bernoulli trial with an appropriate threshold value of $\bm{\tau}$.
This ensures that the probability of selecting slightly more than
$c_k$ tags is sufficiently high while maintaining almost negligible
execution time costs. Please note that, as the Bernoulli trial operates in
parallel on all tags, the communication overhead of OPT-C1, involving
only the transmission of a hash seed and a threshold value, remains minimal.}\\
\emph{\textbf{(p-4) Designing a Refined Sampling Protocol for OPT-C2:}
OPT-C2, tasked with the selection of exactly $c_k$ tags and
their assignment of unique reporting orders, presents an intricate challenge.
It requires ensuring that OPT-C2 operates with an execution time close to the
derived lower bound. Achieving this requires
a lot of time and efforts to get a near optimal answer,
as detailed in subsequent sections.}

}
}

\textcolor{black}{In summary, this paper contributes the following:}\\
\noindent$\bullet$ We  obtain a lower bound on the execution time for
the \textcolor{black}{CIS problem}.\\
$\bullet$ We propose
a near-\underline{OPT}imal \underline{C}ategory-information sampling protocol (OPT-C)
that solves the problem with an execution time
close to the lower bound.\\
$\bullet$ We conduct real-world experiments to verify the practicability and validity of OPT-C; we have also
run extensive simulations comparing OPT-C with state-of-the-art protocols,
and the results clearly show that OPT-C is much faster than existing protocols and is close to the lower bound.\\
\textcolor{black}{$\bullet$ We implement the OPT-C protocol
on commercial off-the-shelf (COTS) devices and test it in
a real RFID system alongside other state-of-the-art C1G2-compatible tag sampling protocols.
The experimental results indicate that the actual implementation of OPT-L (denoted by
$\text{OPT-C}_\mathbf{IMPL}$)
achieves an average reduction in execution time of $37\%$ compared to other protocols.
}

The subsequent sections of this paper are organized as follows:
Section \ref{lb_analysis}
\textcolor{black}{defines}
the \textcolor{black}{CIS problem}
and presents the lower bound
on execution time. Section \ref{protocol} delves into the OPT-C protocol
and establishes its near-optimality. 
\textcolor{black}{Section \ref{sec_imp} focuses on the real-world implementation of OPT-C in RFID systems.
Section \ref{simulation_results} presents the simulation results for OPT-C, showcasing its superior performance.
In Section \ref{experimental_results}, we present experimental findings that highlight the practicality and efficiency of OPT-C's implementation.}
Finally,
Section \ref{conclusion} offers concluding remarks.

\section{Lower bound on execution time}\label{lb_analysis}
This section first defines the Category Information Sampling Problem and then obtains a lower bound on the
execution time required by any protocol that can solve the problem.
\textcolor{black}{To arrive at this lower bound, we employ encoding arguments, a technique commonly used to analyze space constraints in data structures \cite{goswami2014approximate}, along with convex optimization techniques \cite{cambini2008generalized}. Our analysis primarily focuses on the communication exchanges between the reader and the tags when dealing with different input scenarios. The ultimate goal is to determine the minimum number of communication messages needed to effectively handle all possible instances.
}


\subsection{System model and problem formulation}
We focus on a typical RFID system
\cite{
chu2021efficient,liu2022revisiting,zhu2020collisions,9082674,8948356,wzg2019optg}
that consists of a reader $R$
and a population $S$ of $N$ tags.
Reader $R$ connects to the backend server via a high-speed link to obtain computing and storage resources, but it can only communicate with tags via a low-speed wireless link.
A tag in population $S$ has an ID ($96$-bit ID),
which is a unique identifier for both itself and the associated physical object,
and a category-ID, which indicates which category it belongs to.
Tags with the same category-ID carry the same category information.
In short, each tag $t\in S$ has
a unique ID (also represented by $t$), 
a category-ID $t^{\textbf{G}}$,
and a piece of category information.
According to category-IDs,\footnote{\textcolor{black}{In RFID applications, tags are typically categorized based on various criteria, such as shipping destinations or storage locations. Users have the flexibility to customize the category-IDs of tags and pre-write these category-IDs into the User memory of the tags \cite{EPCGlobal}. Subsequently, the reader can query a specific category-ID, denoted as $k$, to select all tags with the corresponding category-ID $k$. This customization allows for a flexible and application-specific categorization approach, which serves as the initial condition for the
{CIS problem}
studied in this paper.
}} population $S$
is partitioned into $K$ categories ($K$ disjoint subsets): $P_1,P_2,...,P_K$,
where each category $P_k$ consists of tags having category-ID $k$, $k\in\{1,2,...,K\}$.
Moreover, let $n_k$ denote the size of category $P_k$, i.e., $n_k$ is the number of tags in $P_k$,
and let $c_k$ denote the reliability number pre-specified by the user for category $P_k$,
i.e., $c_k$ is the number of tags that should be randomly selected from category $P_k$.
\textcolor{black}{Tab. \ref{tab:notations} lists the symbols used in this paper.}

{Initially, reader $R$
knows the IDs and category-IDs of all tags in $S$,
but $R$ has no knowledge of their category information or which of them are currently missing.
Reader $R$ wishes to randomly select $c_k$ tags from each category $P_k$, $k\in\{1,2,...,K\}$,
and collect their category information.\footnote{\textcolor{black}{In this manuscript, we assume that each tag's
category-ID is stored as a substring in the User memory
bank of the tag \cite{EPCGlobal}.
The required length of this substring
to represent $K$ different categories is
$\lceil(\log_2(K)\rceil)$ bits.
Additionally, we assume that tags within a
specific category, denoted as $P_k$, are pre-written
with an identical substring corresponding to the decimal
value of $\mlq \mlq k\mrq\mrq$.
This pre-written substring uniquely identifies the category to which the tags belong.
To solve the
CIS problem,
we use a pre-query mechanism.
Here, the reader initiates a
query for a specific category-ID $k$ to select all tags within category $P_k$.
Following this pre-query, reader $R$ is responsible
for randomly selecting $c_k$ tags from category $P_k$
and assigning unique reporting orders to these selected tags.
These reporting orders are used for collecting category data from each selected tag,
as tags can only be read sequentially one by one.
}}
}

\setcounter{footnote}{0}

\newcommand{\tabincell}[2]{
\begin{tabular}{@{}#1@{}}#2\end{tabular}
}
\begin{table}[!t]
  \centering
  \footnotesize
  \caption{Symbol Definitions}
  \label{tab:notations}
  \begin{tabular}{l|p{27em}}
    \toprule[1pt]
    $U$    & The set of all possible tags with ID values ranging from $0$
    to $\Upsilon-1$, where $\Upsilon=2^{96}$.\tablefootnote{\textcolor{black}{Note that, $U$ also represents
    the set of all possible IDs, i.e., $U=\Big \{ \mathop {000 \cdots 00}\limits_{|\leftarrow\text{96-bit}\rightarrow|},
\mathop {000 \cdots 01}\limits_{|\leftarrow\text{96-bit}\rightarrow|},
\cdot\cdot\cdot\cdot
\mathop {111 \cdots 11}\limits_{|\leftarrow\text{96-bit}\rightarrow|} \Big \}$.}}\\
    \hline
    $S$          &  A tag population of $N$ tags from $U$.\\
    \hline

    $t$ &  A tag in $S$ which has a unique $96$-bit ID (also denoted by $t$), a category-ID $t^{\textbf{G}}\in\{1,2,...,K\}$, and a piece of category information.\\ 
    \hline
    $K$ & The number of categories in $S$.\\
    \hline
    $P_k$ &  The $k$-th category of tags in $S$ ($P_k \subset S$) in which all tags share the same category-ID $k$
    and are written with the same category information related to $P_k$.\\


    \hline
    $c_k$    & The reliability number used for category $P_k$, \textcolor{black}{representing the user wants to select $c_k$ tags
    randomly from $P_k$.}
    \\
    \hline
    $n_k$ & The size of category $P_k$, i.e., the number of tags in $P_k$.\\
    \hline
    $\mathbb{P}$ &  A partition over the universe $U$ according to the written category information on tags,
    and $\mathbb{P}=\{P_1,P_2,...,P_K,U-S\}$ $\forall i\neq j$, $P_i\cap P_j=\emptyset$,
    $\bigcup\nolimits_{k = 1}^K {{P_{\textcolor{black}{k}}}}=S$. Note that $U-S$ is the remainder set containing all tags that do not appear in $S$.
    \\
    \hline
    $R$          &  A reader, who initially knows only the ID and category-ID
    of each tag $t$ in $S$ and wants to select a few random tags from each category $P_k$ and
    \textcolor{black}{assign
    them unique reporting orders.}\\
    \bottomrule[1pt]
  \end{tabular}
\end{table}

\begin{figure}[h]
\centering
\includegraphics[height=3.5cm]{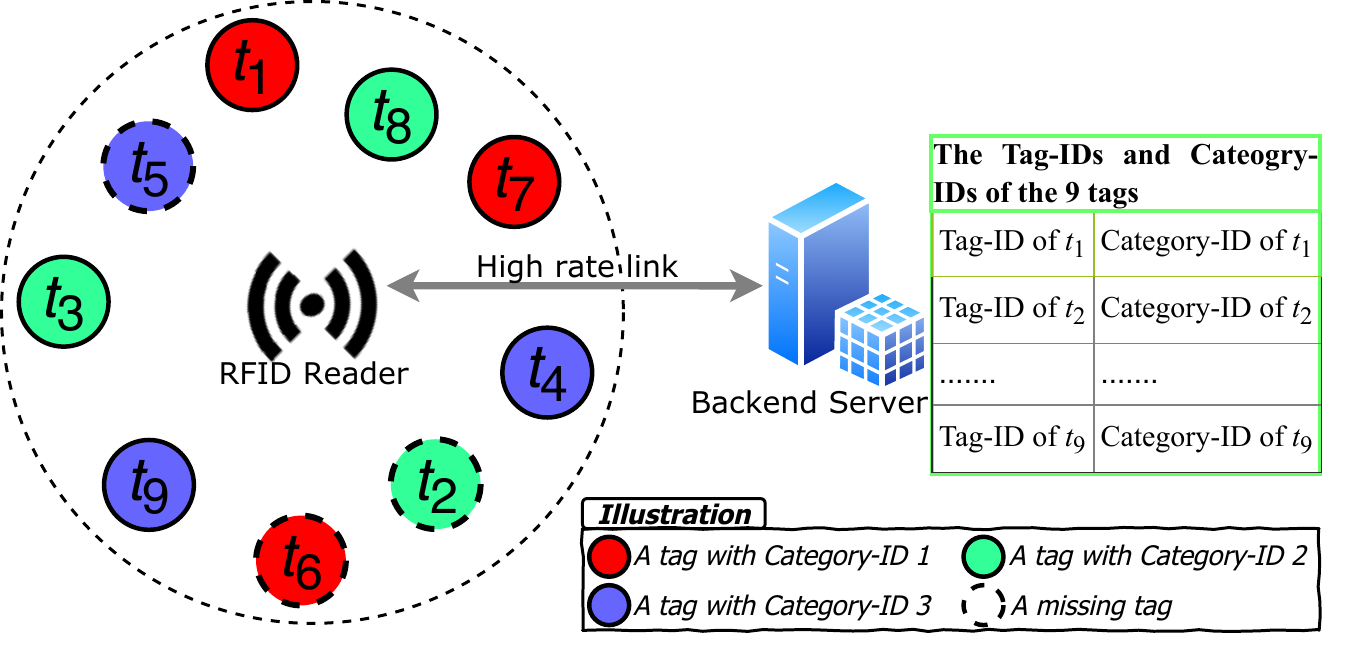}
\caption{An example of the Category Information Sampling Problem in an RFID system with missing tags.
{\footnotesize
Suppose $c_1 =1$, $c_2 = 1$, $c_3=2$. Then
the Category Information Sampling Problem
requires the reader to
randomly select one tag from $P_1$ and read it, randomly select one tag from $P_2$ and read it, and randomly select two tags from $P_3$ and read them.
Note that the three tags: $t_2,t_5,t_6$ are missing, but the reader does not know this fact.
}}
\label{fig-sec}
\end{figure}

\begin{definition}\label{definition_for_category_information_sampling}
Given an RFID system as described above,
the Category Information Sampling Problem
\textcolor{black}{(the CIS problem)}
is to design a protocol $A$ between the reader $R$ and the tags
such that the following two objectives are achieved on each category $P_k$, $k\in\{1,2,\cdots,K\}$.\\
(\textbf{\Rmnum{1}})
{Every subset of $c_k$ tags from $P_k$
has the same probability $1/(_{\,c_k}^{n_k})$ of being
chosen to become the sampled subset $T_k$;}\\
(\textbf{\Rmnum{2}})
Each tag in the sampled subset $T_k$ is informed of a
unique integer from $\{1,2,\cdots,c_k\}$, known as the reporting order of the tag.
\end{definition}
\textcolor{black}{Note:}
the second objective (\textbf{\Rmnum{2}}) is necessary
for collecting category  information from the $c_k$ tags in $T_k$
because reader $R$ can only read the data on the tags sequentially one by one.
\textcolor{black}{Fig. \ref{fig-sec} provides an example of this problem.}

\subsection{A lower bound on execution time}\label{lower_bound_sec}
\textcolor{black}{The following theorem establishes a lower bound
on the communication time between the reader $R$ and tags,
and then obtains a lower bound on the execution time, ${\rm{T}_{{\rm{lb}}}}$.}

First, we present some notations needed in the proof.
Let $U$ represent the universe of all possible tags with ID
values ranging from $0$ up to $\Upsilon-1$, where $\Upsilon=2^{96}$;
As each tag owns a unique ID, $U$ also represents the universe of all $2^{96}$ possible IDs.
Let us call $U-S$ (i.e., $U-\bigcup\nolimits_{k=1}^{K} {P_k}$) the remainder set,
and consider the set-family $\mathbb{P}=\{P_1,P_2,\cdots, P_K, U-S\}$.
Given a tag population size $N$ and $K$ category sizes: $n_1,n_2,\cdots, n_K$,
the universe $U$ can be arbitrarily partitioned into category $P_1$
of $n_1$ tags, category $P_2$ of $n_2$ tags, $\cdots$,
category $P_K$ of $n_K$ tags,
and the remainder set of $\Upsilon-N$ tags,
denoted by $(_{n_1,n_2,\cdots,n_K,\Upsilon-N}^{\quad\;\;\;\;\;\;\;\; \Upsilon})$, i.e.,
partition $\mathbb{P}$ has $(_{n_1,n_2,\cdots,n_K,\Upsilon-N}^{\quad\;\;\;\;\;\;\;\; \Upsilon})$
different possibilities.
In the following,
we use $(_{n_1,n_2,\cdots,n_K}^{\quad\;\;\;\; \Upsilon})$ to
represent the number of possible partitions over the universe $U$,
because
$\sum\nolimits_{k=1}^K n_k = N$ and
$(_{n_1,n_2,\cdots,n_K,\Upsilon-N}^{\quad\;\;\;\;\;\;\;\; \Upsilon})=
(_{n_1,n_2,\cdots,n_K}^{\quad\;\;\;\; \Upsilon})$.\footnote{$\sum\nolimits_{k=1}^K n_k = N$ comes from the fact that a tag in $S$ belongs to one category and $\bigcup\nolimits_{k = 1}^K {{P_k}}=S$.
}

When exploring the lower bound of protocol execution time,
we take the communication time between reader $R$ and tags as the primary determinant,
since $R$ has a high-speed connection to the backend server but communicates with tags via a low-speed link slowly
\cite{liu2018efficient,liu2022revisiting,qiao2015tag,wzg2019optg}.
In addition, we omit the computation time of reader $R$ since the backend server can provide computation assistance very quickly, and our lower bound is still valid even with computation time included.




\begin{theorem}\label{lowerbound}
Any protocol $A$ that solves the
Category Information Sampling Problem described
in Definition \ref{definition_for_category_information_sampling},
must satisfy the following two inequalities
\begin{flalign}
&|A| \ge  \Gamma =  \sum\nolimits_{k=1}^K{\log_2(e)c_{\color{black}k}}
\label{lb_for_perfect_channel_inequ},\\
&{\rm{T}_{{\rm{lb}}}} \ge {{\rm{T}}_{{\rm{96}}}}\times\Gamma/96 \label{lb_for_perfect_channel_time_lb},
\end{flalign}
where ${\rm{T}}_{{\rm{96}}}$ is the time cost
of transmitting \textcolor{black}{$96$ bits between reader $R$ and tags},
$|A|$ represents the number of bits communicated between reader $R$ and tags during $A$'s execution,
and ${\rm{T}_{{\rm{lb}}}}$ represents the lower bound of $A$'s execution time.
\begin{IEEEproof}
Let $\textit{M}$ denote
the complete set of communicated messages between reader $R$ and tags
during $A$'s execution.
Then, $|A|$
equals the number of bits contained in $\textit{M}$.
Clearly, if protocol $A$ can solve the
\textcolor{black}{CIS problem}
with $K$ given category sizes: $n_1, n_2,...,n_K$, 
it must take care of all the $(_{n_1,n_2,\cdots,n_K}^{\quad\;\;\;\; \Upsilon})$
different partitions.
Therefore, in order to obtain the required lower bound, we will
study how many bits are needed for message $\textit{M}$ to handle all these partitions.
Below is a roadmap of our proof.
{\color{black}{
\begin{itemize}
\item First, we model the decision process on each tag $t\in S$
and explain how message $\textit{M}$
helps $t$ to decide whether it is sampled or not,
and if so, what its reporting order is.
\item Second, we analyze how a specific value $v$ of message $\textit{M}$
can help solve the
\textcolor{black}{CIS problem}
and how many different partitions it can handle;
\item Third, since there are $(_{n_1,n_2,..,n_K}^{\quad\;\;\;\; \Upsilon})$ different partitions to be handled when solving this problem,
we can obtain the lower bound on $\textit{M}$
by analyzing how many different values the message $\textit{M}$ must have to handle all these partitions.
\end{itemize}
}}

As a starting point, we model each tag's decision process and describe
how each tag changes its state after protocol A's execution
(i.e., after receiving message $\textit{M}$).
In this way, we lay the foundation for our proof.
Without loss of generality, we denote the internal decision process
on each tag $t\in S$ by a function $F()$.
Initially, each tag $t\in S$ is in the same state,
{which means it doesn't know whether it needs to report or not, and if so, what its report number is.}
After receiving message $\textit{M}$, tag $t$ takes
a category-ID $k$ ($k\in\{1,2,...,K\}$), $\textit{M}$, and
its ID (also represented by $t$) as the three input parameters
to function $F()$, and then computes
$F(k,\textit{M},t)$ to decide whether
{it will be a sampled tag in category $P_k$ (the $k$-th category), and if so,}
what its reporting order is.
More specifically, without sacrificing generality,
we assume the following decision rules on tag $t$.
\begin{itemize}
\item[$\blacktriangleright$] If $F(k,\textit{M},t)=0$,
\textcolor{black}{tag} $t$
identifies itself as \textcolor{black}{a non-sampled tag}
in category $P_k$.
\item[$\blacktriangleright$] If $F(k,\textit{M},t)=i\in\{1,2,\cdots,c_k\}$,
tag
$t$ identifies itself as a sampled tag in category $P_k$ and
takes $i$ as its reporting order.
\end{itemize}
Clearly, message $\textit{M}$ is the parameter that determines
the behavior of the $K$ functions:
$F(k,\textit{M},\bullet)$, $k=1,2,...,K$.
That is, a specific value $v$ of $\textit{M}$ ($\textit{M}$ is used as the second parameter of function $F()$)
uniquely determines how function $F(k,\textit{M},\bullet)$
maps the $\Upsilon = 2^{96}$ tags from $U$ into the range $\{0,1,2,...,c_k\}$.

Next, we analyze the number of different partitions that a specific value \textcolor{black}{$v$} can handle.
Let $S^v_{k,i}$ represent the set consisting of
those tags from $U$ that have function $F(k,v,\bullet)$ equal to $i$, or namely the following:
\begin{flalign}\label{definition_of_function_value_subset}
&S^v_{k,i}=\{t|t\in U \,\text{and}\, F(k,v,t)=i\},\nonumber\\
&\quad\quad\quad\quad\quad\quad k\in\{1,2,...,K\},i\in\{1,2,...,c_k\}.
\end{flalign}
Then, $v$ can handle any partition $\mathbb{P}=\{P_1,P_2,...,P_K\}$ that is constructed by (s1)-(s5):
\begin{itemize}[leftmargin=0.7cm]
\item[(s1)] Choose one tag from each of the $c_1$ sets: $S^{v}_{1,1},S^{v}_{1,2},\cdots, S^{v}_{1,c_1}$,
choose $n_1-c_1$ tags from $S^{v}_{1,0}$, and group these $n_1$ tags
into the first category $P_1$.
\item[(s2)] Delete the chosen $n_1$ tags from the $c_2+1$ sets: $S^{v}_{2,0},S^{v}_{2,1},\cdots, S^{v}_{2,c_2}$.
\emph{Note that \textcolor{black}{(S1) has put}
these $n_1$ tags into the $c_1+1$ sets:
    $S^{v}_{1,0},S^{v}_{1,1},\cdots, S^{v}_{1,c_1}$.}
\item[(s3)] Choose one tag from each of the $c_2$ sets: $S^{v}_{2,1},S^{v}_{2,2},\cdots, S^{v}_{2,c_2}$,
choose $n_2-c_2$ tags from $S^{v}_{2,0}$, and group these $n_2$ tags
into the second category $P_2$.
\item[(s4)] Delete the chosen $n_1+n_2$ tags from the $c_3+1$ sets: $S^{v}_{3,0},S^{v}_{3,1},\cdots, S^{v}_{3,c_3}$.
\emph{Note: (s1) and (s2) have put
these $n_1+n_2$ tags into the $c_1+c_2+2$ sets:
    $S^{v}_{1,0},S^{v}_{1,1},\cdots, S^{v}_{1,c_1},S^{v}_{2,0},S^{v}_{2,1},\cdots, S^{v}_{2,c_2}$.
    }
\item[(s5)] Repeat the same process for
\textcolor{black}{$P_3, P_4, \cdots, P_K$.}
\end{itemize}
By the above (s1)-(s5) and the definition of $S^v_{k,i}$  in \eqref{definition_of_function_value_subset},
it is clear that,
the tag chosen from $S^v_{k,1}$ will be the sampled tag with reporting order $1$,
the tag chosen from $S^v_{k,2}$ will be the sampled tag with reporting order $2$, $\cdots$,
and the tag chosen from $S^v_{k,c_k}$ will be the sampled tag with reporting order $c_k$.
These $c_k$ tags together form the sample subset $T_k$
for category $P_k$, while the $n_k-c_k$ tags chosen from $S^v_{k,0}$ are the unsampled tags in $P_k$.\footnote{
\textcolor{black}{Let us explain (s1).}
Assume that $t_1$
is the tag chosen from $S^v_{1,1}$, $t_2$ is the tag chosen from
$S^v_{1,2}$, $\cdots$, $t_{c_1}$ is the tag chosen from $S^v_{1,c_1}$,
and $t_{c_1 +1}, t_{c_1 +2},...,t_{n_1}$ are the $n_1 - c_1$ tags chosen from $S^v_{1,0}$.
$t_1$ will be the sampled tag with reporting order $1$ in $P_1$ since $F(k=1,v,t_1)=1$;
$t_2$ will be the sampled tag with reporting order $2$ in $P_1$ since $F(k=1,v,t_2)=2$;
$\cdots$;
$t_{c_1}$ will be the sampled tag with reporting order $c_1$ in $P_1$ since $F(k=1,v,t_{c_1})=c_1$.
These $c_1$ tags
together
form the sample subset $T_1$
for $P_1$.
The $n_1-c_1$ tags: $t_{c_1 + 1}, t_{c_1 +2},...,t_{n_1}$ chosen from $S^v_{1,0}$ will be
unsampled tags in $P_1$ since
$F(k=1,v,t_{c_{1} +1})=F(k=1,v,t_{c_{1} +2})=...=F(k=1,v,t_{n_1})=0$.}
Therefore, $v$ achieves the objectives required by the
\textcolor{black}{CIS problem}
(see Definition \ref{definition_for_category_information_sampling}),
and
we can say that $v$ is able to handle all the partitions
created by this five-step process (s1)-(s5).


\begin{table}[!htb]
\caption{The values of two functions: $F(1,v,\bullet)$
and $F(2,v,\bullet)$
 over each tag in the universe $U=\{\texttt{0},\texttt{1},...,\texttt{6}\}$.}\label{fun_value_table}
 \centering
\begin{tabular}{|r!{\vrule width 1pt}*{7}{c|}}\hline
\backslashbox{~}{$t\in U$~~}
  & \texttt{0}& \texttt{1} & \texttt{2} & \texttt{3} & \texttt{4} &\texttt{5} &\texttt{6} \\\noalign{\hrule height 0.75pt}
Function $F(1,v,\bullet)$ & 0& 0& 0& 0& 1& 1& 1\\\hline
Function $F(2,v,\bullet)$   &0&0&0&0&0&1&1\\\hline
\end{tabular}
\end{table}

Let us use an example to illustrate how a value $v$ of $\textit{M}$
handles multiple partitions.
In this example, 
we assume the universe $U$ contains $7$ tags, $U=\{\texttt{0},\texttt{1},...,\texttt{6}\}$,
i.e., each tag has a unique ID from $\texttt{0}$ to $\texttt{6}$.
The tag population $S$ contains $N=5$ tags of $U$,
and $S$ is divided into two subsets (two categories) $P_1$ and $P_2$,
with $|P_1|=3$ and $|P_2|=2$.
We consider the
\textcolor{black}{CIS problem}
that requires
randomly picking $c_1=1$ tag from $P_1$ and $c_2=1$ tag from $P_2$.
Moreover, suppose that a specific value $v$ of
message $\textit{M}$ uniquely determines
two specific functions, as shown in Tab. \ref{fun_value_table}.
By this table, we have $S^v_{1,0}=\{\texttt{0},\texttt{1},\texttt{2},\texttt{3}\}$, $S^v_{1,1}=\{\texttt{4},\texttt{5},\texttt{6}\}$,
$S^v_{2,0}=\{\texttt{0},\texttt{1},\texttt{2},\texttt{3},\texttt{4}\}$,
and $S^v_{2,1}=\{\texttt{5},\texttt{6}\}$.
Then,  $v$ can handle the partition $\mathbb{P}=\{P_1,P_2\}=\{\{\texttt{0},\texttt{3},\texttt{5}\},\{\texttt{1},\texttt{6}\}\}$,
because only tag $\texttt{5}$ is picked from $P_1$ and is assigned reporting order $1$ ($F(1,v,\texttt{5})=1$),
and only tag $\texttt{6}$ is picked from $P_2$ and is assigned reporting order $1$ ($F(2,v,\texttt{6})=1$).
\textcolor{black}{Tags} $\texttt{0}$ and $\texttt{3}$ in $P_1$ are not chosen because $F(1,v,\texttt{0})=F(1,v,\texttt{3})=0$;
tag $\texttt{1}$ in $P_2$ is not chosen because $F(2,v,\texttt{1})=0$.
Besides,
$v$ can 
solve any partition $\mathbb{P}$ constructed as below:

\indent (1) Choose one tag from $S^v_{1,1}$, two tags from $S^v_{1,0}$,
and group these three tags into category $P_1$;

\indent (2) Choose one tag from $S^v_{2,1}-P_1$, one tag from $S^v_{2,0}-P_1$,
and group these two tags into category $P_2$.



In general,
let $s^v_{k,i}=|S^v_{k,i}|$ ($k\in\{1,2,..,K\}$, $i\in \{0,1,..,c_k\}$),
and we can find the following fact for the
\textcolor{black}{CIS problem}
with
$K$ category sizes:
$n_1,n_2,..,n_K$
and $K$ reliability numbers: $c_1,c_2,...,c_K$.\\
\emph{
$\star$ \textbf{FACT-1}:
There are $(_{n_k-c_k}^{\;\;s^v_{k,0}})\prod\nolimits_{i=1}^{c_k}s^v_{k,i}$
different ways to choose tags from $S^v_{k,0},S^v_{k,1},...,S^v_{k,c_k}$
to create a category $P_k$ that can be handled by value $v$;
and $s^v_{k,0},s^v_{k,1},...,s^v_{k,c_k}$
satisfy $\sum\nolimits_{i=0}^{c_k}s^v_{k,i}=|U-\bigcup\nolimits_{\textcolor{black}{j}=1}^{k-1} {P_i}|=\Upsilon-\sum\nolimits_{j=1}^{k-1}n_j$.
}\\
Let us first see why \textbf{FACT-1} is true for category $P_1$.
Since all tags in $U$ have not been chosen yet,
we can choose exactly one tag from each of the $c_1$ sets: $S^{v}_{1,1}, S^{v}_{1,2},..., S^{v}_{1,c_1}$,
and choose $n_1-c_1$ tags from $S^v_{1,0}$ to create $P_1$.
So, there are $(_{n_1-c_1}^{\;\;s^v_{1,0}})\prod\nolimits_{i=1}^{c_1}s^v_{1,i}$ different
ways to create $P_1$ that can be handled by $v$,
and $s^v_{1,0},s^v_{1,1},...,s^v_{1,c_1}$ satisfy $\sum\nolimits_{i=0}^{c_1}s^v_{1,i}=\Upsilon$.
Following a similar analysis, we can show that this fact is also true for
$P_2$, $P_3$, $\cdots$, $P_K$.


With \textbf{FACT-1}, the number of partitions that $v$ can handle is
\begin{equation}\label{number_of_partitions_handled_by_v1}
\prod\nolimits_{k=1}^{K}\left((_{n_k-c_k}^{\;\;s^v_{k,0}})\prod\nolimits_{i=1}^{c_k}s^v_{k,i}\right),
\end{equation}
which can \textcolor{black}{be} proven to be less than or equal to
\begin{equation}\label{number_of_partitions_handled_by_v2}
\prod\nolimits_{k=1}^{K} \left[\frac{(n_k-c_k)^{n_k-c_k}}{(n_k-c_k)!}\left( \frac{\Upsilon-\sum\nolimits_{j=1}^{k-1}n_j}{n_k}\right)^{n_k}\right]=\Phi.\footnote{
\textcolor{black}{A
detailed proof of the inequality
$\eqref{number_of_partitions_handled_by_v1} \le \eqref{number_of_partitions_handled_by_v2}$
is in Appendix B.}}
\end{equation}


Finally, because there are $(_{n_1,n_2,..,n_K}^{\quad\;\;\;\; N})$ different partitions to process,
and a specific value $v$ of message $\textit{M}$ can handle at most $\Phi$ of them,
\textcolor{black}{we know that message $\textit{M}$ needs to have at least
$(_{n_1,n_2,..,n_K}^{\quad\;\;\;\; N})/{\Phi}$ different values
\cite{goswami2014approximate}}.
Let's further analyze $(_{n_1,n_2,..,n_K}^{\quad\;\;\;\; N})/{\Phi}$ as follows:
{\small{
\begin{flalign}
&{(_{n_1,n_2,..,n_K}^{\quad\;\;\;\; N})}/{\Phi} = {
\left(\prod\nolimits_{k=1}^{K}\big(_{\quad\quad n_k}^{\Upsilon-\sum\nolimits_{j=1}^{k-1}n_j}\big)
\right)}/{{\Phi}}\nonumber\\
&=\prod\limits_{k=1}^{K} \frac{\big(_{\quad\quad n_k}^{\Upsilon-\sum\nolimits_{j=1}^{k-1}n_j}\big)}{ \frac{(n_k-c_k)^{n_k-c_k}}{(n_k-c_k)!}\big( \frac{\Upsilon-\sum\nolimits_{j=1}^{k-1}n_j}{n_k}\big)^{n_k}}\approx\prod\limits_{k=1}^K\frac{ {(n_k-c_k)!} {n_k}^{n_k} }{n_k! (n_k-c_k)^{n_k-c_k}}\nonumber\\
&\ge \prod\nolimits_{k=1}^K \frac{ \sqrt{2\pi} (n_k-c_k)^{n_k-c_k+0.5} e^{-(n_k-c_k)} {n_k}^{n_k}  }
{ e {n_k}^{n_k+0.5} e^{-n_k} (n_k-c_k)^{n_k-c_k} }\nonumber \\
&= \prod\nolimits_{k=1}^K (\sqrt{2\pi}/e) ((n_k-c_k)/n_k)^{0.5} e^{c_k}. \label{number_of_value_message}
\end{flalign}
}}
In the above,
the approximation  in the second line comes from
$(\Upsilon - \sum\nolimits_{j=1}^{k-1} n_j)^{n_k}\approx (\Upsilon - \sum\nolimits_{j=1}^{k-1} n_j)^{\underline{n_k}}$,
which is true
because $\Upsilon \gg \sum\nolimits_{j=1}^K n_j=N$
{(A real RFID system usually contains less than $10^8$ tags, i.e., $N\le10^8$, so $N\ll\Upsilon=2^{96}\approx 8\times 10^{28}$);}
the inequality in the third line comes from $(n_k-c_k)!\ge \sqrt{2\pi} (n_k-c_k)^{n_k-c_k+0.5} e^{-(n_k-c_k)}$
and ${n_k}!\le e {n_k}^{n_k+0.5} e^{-n_k}$, which is based on Stirling's Formula (see Lemma 7.3 in \cite{mitzenmacher2017probability}).

Taking the logarithm of \eqref{number_of_value_message}, we can
find out that
message $\textit{M}$ must contain $\log_2 (\prod\nolimits_{k=1}^K (\sqrt{2\pi}/e) ((n_k-c_k)/n_k)^{0.5} e^{c_k}) \approx
\sum\nolimits_{k=1}^K{\log_2(e)c_{\textcolor{black}{k}}}$ bits. This leads to
the lower bound ${\rm{T}_{{\rm{lb}}}}$ on the execution time of protocol $A$.\footnote{\textcolor{black}{Hence, the lower bound is mainly related to $K$ and the $K$ reliability
numbers $c_1, c_2, ..., c_K$, as it increases when $c_1, c_2, ..., c_K$ and $K$
increase. However, it remains relatively independent of the change of $n_1, n_2, ..., n_K$.
A detailed explanation is given in Appendix F.}}

\end{IEEEproof}
\end{theorem}


\section{A near-optimal protocol for the Category Information Sampling Problem (OPT-C)}\label{protocol}
\textcolor{black}{Motivated by the lower bound established in Theorem \ref{lowerbound} we introduce OPT-C, a near-optimal category information sampling protocol with the capacity to approach this lower bound. Tab. \ref{tab:th_cmp} provides a comparative illustration of OPT-C's theoretical superiority over existing state-of-the-art sampling protocols.}

{\color{black}{
Generally, OPT-C's design revolves around two fundamental stages:
In the initial stage, OPT-C employs a coarse sampling process to select approximately $c_k$ tags from each category $P_k$. This process creates a Bernoulli sample from the tags within $P_k$, with tags chosen independently in parallel. However, the drawback of this method is that the resulting sample size fluctuates as it becomes a random variable, rather than a fixed quantity.
To mitigate this variability and ensure selection of precisely $c_k$ tags from $P_k$,
OPT-C introduces a second stage, comprising a refined sampling process. In this phase, exactly $c_k$ tags are selected one by one from the Bernoulli sample generated in the first stage. This process results in a simple random sample from category $P_k$, with each selected tag assigned a unique reporting order.

The core design concept underpinning OPT-C can be summarized as follows:
\emph{On one hand, directly creating a random sample containing precisely $c_k$ tags from category $P_k$ would effectively resolve
\textcolor{black}{the CIS problem.}
However, this approach proves time-consuming when applied to categories with a large number of tags. The collective decision-making required among all tags within $P_k$ is the reason for the time inefficiency, as it ensures that exactly $c_k$ of them are sequentially included in the sample.
On the other hand, a Bernoulli sample is created quickly by independently conducting Bernoulli trials on all tags within $P_k$ in parallel. However, the downside is the inherent variability in the number of tags within this sample, rendering it a random variable rather than a predetermined constant.}
Hence, within OPT-C, we adopt a hybrid approach. Firstly, we design a coarse sampling protocol that
promptly generates a random sample containing slightly more than $c_k$ tags from $P_k$ using Bernoulli trials. Subsequently, we introduce a refined sampling protocol that randomly selects exactly $c_k$ tags from this previously generated sample, one after another. Importantly, this refined sampling process incurs a time cost that approaches the lower bound, ensuring efficient execution.
}}

\textcolor{black}{T}o distinguish the tags selected by OPT-C1 and those further chosen by OPT-C2 for reporting, each tag $t$ in the population $S$ assumes one of the following four statuses:\\
$-$\textbf{Unacknowledged state}: Every tag $t\in S$ is initially in this state, signifying that it has not yet decided whether to report or not.\\
$-$\textbf{Unselected state}: A tag $t$ enters this state when it decides not to report its category information.\\
$-$\textbf{Selected state}: A tag $t$ enters this state when it chooses to report its category information but has not yet received a unique reporting order.\\
$-$\textbf{Ready state}: A tag $t$ enters this state when it has decided to report its category information and has been assigned a unique reporting order.

\begin{table}[!t]
  \centering
  \footnotesize
  \caption{Theoretical comparisons of existing category information sampling protocols}
  \label{tab:th_cmp}
  \begin{tabular}{|p{3em}!{\vrule width 1pt}p{10em}|p{14em}|}
    \toprule[1.2pt]
    Protocol    & Total execution time & Requirements on tags\\
    \midrule[1.1pt]
    TPS \cite{liu2018efficient}&  unknown
    (no closed form expression for the total execution time) & \emph{Each tag supports
    \textbf{(1)}: a hash function  and \textbf{(2)}: the operation that counts the number `$1$'s in a bit-array.}
    \\\hline
    ACS \cite{chu2021efficient}& unknown (no closed form expression for the total execution time) &  \emph{Each tag supports
    \textbf{(1)}: a hash function, \textbf{(2)}: the operation that counts the number $\mlq 1 \mrq$s in a bit-array,
    and \textbf{(3)} the arithmetic coding algorithm that decompresses received frames.}
    \\\\\hline
    OPT-C (our solution) & $\sum\nolimits_{k=1}^K \big[{e c_k}+ \log_2{(\frac{96}{\log_2n_k})} +\log_2 c_k\big]$& \emph{Each tag supports
    \textbf{(1)}: a hash function and \textbf{(2)}: the operation that counts the number `$1$'s in a bit-array.}
    \\
    \midrule[1.1pt]
    The lower bound
     &
     \multicolumn{2}{|p{24em}|}{
     $\sum\nolimits_{k=1}^K{\log_2(e)c_{\textcolor{black}{k}}}$
     (The theoretical performance limit for any category information sampling protocol)}\\
    \bottomrule[1.2pt]
  \end{tabular}
\end{table}

\subsection{The design and analysis of OPT-C1}\label{opt-c1}
OPT-C1 contains $K$ communication rounds: $\mathbb{I}_1, \mathbb{I}_2,\cdots,\mathbb{I}_K$, where
each round $\mathbb{I}_k$, \textcolor{black}{$k=1,2,\cdots,K$},
generates a random sample of a little more than $c_k$ tags from category $P_k$. 
In each round $\mathbb{I}_k$, reader $R$ sends a random seed to all tags in $P_k$,
and then each tag in $P_k$ performs a Bernoulli trial
to determine whether to enter the selected state or the unselected state.
The detailed steps are given below.\\
{\indent\textbf{Step1:} Reader $R$ broadcasts
two parameters $<r,c_k>$ out to all tags,
where $r$ is a suitable
random seed chosen by reader $R$,} and $c_k$ is the reliability number.\\
\indent\textbf{Step2:} Upon receiving $<r,c_k>$, each tag $t$ in category $P_k$
computes function $h(t)=H(t,r)\bmod n_k$, and
if $h(t) \le c_k + 3\sqrt{c_k}$, tag $t$ changes its status from the unacknowledged
state to selected state; otherwise, $t$ enters the unselected state.

In Step1, we call a random seed $r$ \textbf{suitable}
in the sense that, at least $c_k$ and at most $c_k+6\sqrt{c_k}$ tags from
$P_k$ change their status from the unacknowledged state to the selected state when
they use $r$.
We claim that reader $R$ can always find a suitable random seed $r$,
because: (\rmnum{1})
{Given any random seed $r$, the probability that it is a suitable random seed is higher than $0.4$
(see Theorem \ref{pick_probability_OPT-C1});}
and (\rmnum{2}) with a probability almost equal to $1$,
reader $R$ tests no more than $6$ random seeds
until a suitable one is found (Theorem \ref{number_of_trials_4_find_random_seeds}).


{\color{black}
\begin{table}[!htb]
    \caption{The hash values and states of the tags in a category $P_k=\{t_1,t_2,...,t_{12}\}$
    when using random seed $r_1$ (A) and $r_2$ (B) in OPT-C1.}
    \label{hash_tab}
    \begin{minipage}{0.49\linewidth}
      \centering
       (A) Using random seed $r_1$
       \begin{tabular}{|c!{\vrule width 1pt}m{4em}|c|} 
      \hline
      &$h(t)=H(t,r_1)$ mod $12$& \tabincell{l}{State after\\OPT-C1}\\
     \noalign{\hrule height 1pt}
      $t_1$& {\textbf{$\quad\;$11}} & {{unselected}} \\
      \hline
      \tabincell{l}{$t_2$}& {\textbf{$\quad\;$6}} & {{{selected}}}\\
      \hline
      \tabincell{l}{$t_3$}&  {\textbf{$\quad\;$3}} & {{{selected}}}\\
      \hline
      \tabincell{l}{$t_4$}&  {\textbf{$\quad\;$4}} & {{{selected}}}\\
      \hline
      \tabincell{l}{$t_5$}&  {\textbf{$\quad\;$3}} & {{{selected}}}\\
      \hline
      \tabincell{l}{$t_6$}&  {\textbf{$\quad\;$5}} & {{{selected}}}\\
      \hline
      \tabincell{l}{$t_7$}&  {\textbf{$\quad\;$6}} & {{{selected}}}\\
      \hline
      \tabincell{l}{$t_8$}&  {\textbf{$\quad\;$1}} & {{{selected}}}\\
      \hline
      \tabincell{l}{$t_9$}&  {\textbf{$\quad\;$1}} & {{{selected}}}\\
      \hline
      \tabincell{l}{$t_{10}$}&  {\textbf{$\quad\;$0}} & {{{selected}}}\\
      \hline
      \tabincell{l}{$t_{11}$}&  {\textbf{$\quad\;$2}} & {{{selected}}}\\
      \hline
      \tabincell{l}{$t_{12}$}&  {\textbf{$\quad\;$2}} & {{{selected}}}\\
     \hline
    \end{tabular}
    \end{minipage}%
    \begin{minipage}{0.49\linewidth}
      \centering
      (B) Using random seed $r_2$
        \begin{tabular}{|c!{\vrule width 1pt}m{4em}|c|} 
      \hline
      &$h(t)=H(t,r_2)$ mod $12$& \tabincell{l}{State after\\OPT-C1}\\
      \noalign{\hrule height 1pt}
      $t_1$&  {$\quad\;\;$\textbf{0}} & {{{selected}}} \\
      \hline
      \tabincell{l}{$t_2$}&  {$\quad\;\;$\textbf{8}} & {{unselected}}\\
      \hline
      \tabincell{l}{$t_3$}&   {$\quad\;\;$\textbf{2}} & {{{selected}}}\\
      \hline
      \tabincell{l}{$t_4$}&  {$\quad\;\;$\textbf{11}} & {{unselected}}\\
      \hline
      \tabincell{l}{$t_5$}&   {$\quad\;\;$\textbf{3}} & {{{selected}}}\\
      \hline
      \tabincell{l}{$t_6$}&  {$\quad\;\;$\textbf{9}} & {{unselected}}\\
     \hline
     \tabincell{l}{$t_7$}&   {$\quad\;\;$\textbf{4}} & {{{selected}}}\\
      \hline
      \tabincell{l}{$t_8$}&  {$\quad\;\;$\textbf{7}} & {{unselected}}\\
      \hline
      \tabincell{l}{$t_9$}&   {$\quad\;\;$\textbf{6}} & {{{selected}}}\\
      \hline
      \tabincell{l}{$t_{10}$}&   {$\quad\;\;$\textbf{5}} & {{{selected}}}\\
      \hline
      \tabincell{l}{$t_{11}$}&  {$\quad\;\;$\textbf{10}} & {{unselected}}\\
      \hline
      \tabincell{l}{$t_{12}$}&  {$\quad\;\;$\textbf{7}} & {{unselected}}\\
     \hline
    \end{tabular}
    \end{minipage}
\end{table}

Check an example to see how OPT-C1 works for a category $P_k$.
Suppose a category $P_k$ contains $12$ tags: $\{t_1,t_2,...,t_{12}\}$,
and we want to select $c_k=2$ tags randomly from $P_k$.
Now, assume that reader $R$ tries to use a random seed $r_1$ for this job,
and part (A) of Tab. \ref{hash_tab}
gives the hash values of the $12$ tags when using the random seed $r_1$.
Then, by step 2 of OPT-C1, tag $t_1$ enters the unselected state because its hash value $h(t_1)=11$
is larger than $c_k+3\sqrt{c_k}=2+3\sqrt{2}\approx6.24$,
while all the other tags enter the selected state because
their hash values are all less than $6.24$.
So, random seed $r_1$ is not suitable because the number of selected tags is $11$ which is not within the range of $[c_{k},c_{k}+6\sqrt{c_k}]=[2,10]$.
As mentioned before, reader $R$ can predict the result of using random seed $r_1$ since it has the full knowledge of the IDs and category-IDs of tags and can pre-compute tags' hash values when using $r_1$. Therefore, $R$ will not broadcast $r_1$, but will try another random seed $r_2$.
The hash values of the $12$ tags when using $r_2$
are given in part (B) of Tab. \ref{hash_tab}.
Then, by step 2 of OPT-C1, we can observe that
\textcolor{black}{tags}
$t_1,t_3,t_5,t_7,t_9,t_{10}$ enter the selected state because their hash values are less than
$c_k+3\sqrt{c_k}=2+3\sqrt{2}\approx6.24$, while the other $6$ tags enter the unselected states as their
hash values are larger than $6.24$.
Random seed $r_2$ is suitable
because it brings $6$ tags of $P_k$ to the selected state,
and $6\in [c_k,c_k+6\sqrt{c_k}]=[2,10]$.
So, finally, reader $R$ will broadcast $r_2$ to tags.
Sometimes a random seed comes to be suitable, and sometimes it does not.
However, we shall show in Theorem \ref{pick_probability_OPT-C1} that
{the probability of an arbitrary random seed being a suitable random seed for round $\mathbb{I}_k$}
(i.e., for category $P_k$)
is greater than $0.4$.
As a result, on average, the reader $R$ only needs to attempt $2.5$ trials to find a suitable random seed (see Theorem \ref{number_of_trials_4_find_random_seeds}).

\begin{theorem}\label{pick_probability_OPT-C1}
Let $r$ be an arbitrary random seed,
then there is
a probability greater than $0.4$
that
$r$ can bring at least $c_k$ tags and at most $c_k+6\sqrt{c_k}$ tags of
$P_k$ to the selected state.
\begin{IEEEproof}
For illustration, let $t_1,t_2,\cdots, t_{n_k}$ be the $n_k$ tags in $P_k$.
We define $n_k$ 
random variables: $X_1,X_2,\cdots, X_{n_k}$, where
$X_i=1$, $i=1,2,..., n_k$, if tag $t_i$ changes its status from the unacknowledged state
to selected state and $0$ otherwise.
Because $h(t_i) = H(t_i,r)\bmod n_k$ is a hash function that
maps a tag $t_i\in P_k$ ($i=1,2,\cdots, n_k$) uniformly at random to
an integer in $\{1,2,...,n_k\}$\footnote{
Here, for analytical convenience,
we slightly modify the hash function $h(t_i)$ to be
$h(t_i)=\left(H(t_i,r)\bmod n_k\right) +1$,
which changes its value domain from $\{0,1,...,n_k-1\}$ to $\{1,2,...,n_k\}$.
By doing this, we avoid the annoying $0$ in the value domain
and can say $\textbf{Pr}(X_i=1) = \textbf{Pr}(h(t_i) \le c_k + 3\sqrt{c_k})=\frac{c_k+3\sqrt{c_k}}{n_k}$.
}, we can get
$\textbf{Pr}(X_i=1) = \textbf{Pr}(h(t_i) \le c_k + 3\sqrt{c_k})=\frac{c_k+3\sqrt{c_k}}{n_k}$.


Let $X=\sum\nolimits_{i=1}^{n_k} X_i$.
Then, $E[X] = \sum\nolimits_{i=1}^{n_k} 1\times \textbf{Pr}(X_i=1) 
=c_k+3\sqrt{c_k}$.
Since each tag $t_i\in P_k$
computes its hash function $h()$ independently,
we can know that $X_1,X_2,\cdots, X_{n_k}$
are independent random variables.
Then using the Chernoff bounds shown in (4.1) and (4.4) of Ref. \cite{mitzenmacher2017probability},
the following two inequalities are true for any $\delta\in(0,1)$:
\begin{flalign}
&\textbf{Pr}(X > (1+\delta)E[X]) < \left({e^{\delta}}/{(1+\delta)^{1+\delta}}\right)^{E[X]}, \label{probability_upper}\\
&\textbf{Pr}(X < (1-\delta)E[X]) \le \left({e^{-\delta}}/{(1-\delta)^{1-\delta}}\right)^{E[X]}. \label{probability_lower}
\end{flalign}

Next, we put $E[X]=c_k+3\sqrt{c_k}$ and $\delta = \frac{3\sqrt{c_k}}{c_k+3\sqrt{c_k}}$ into
 \eqref{probability_upper}, and then get
{\small{
\begin{flalign}
&\textbf{Pr}(X > (1+\frac{3\sqrt{c_k}}{c_k+3\sqrt{c_k}})(c_k+3\sqrt{c_k})) = \textbf{Pr}(X > c_k+6\sqrt{c_k}) \nonumber\\
&< \left(\frac{e^{{3\sqrt{c_k}}/({c_k+3\sqrt{c_k}})}}{(1+{3\sqrt{c_k}}/({c_k+3\sqrt{c_k}}))^{1+{3\sqrt{c_k}}/({c_k+3\sqrt{c_k}})}}\right)^{c_k+3\sqrt{c_k}}
\nonumber\\
& = \frac{e^{3\sqrt{c_k}}}{(1+\frac{3\sqrt{c_k}}{c_k+3\sqrt{c_k}})^{c_k+6\sqrt{c_k}}} 
 \le \frac{e^{3}}{(1+\frac{3}{1+3})^{1+6}} = 0.3996 \label{probability_upper2}.
\end{flalign}
}}
The inequality in \eqref{probability_upper2} is due to
the fact that $c_k\ge 1$
and function
$e^{3\sqrt{c_k}}(1+\frac{3\sqrt{c_k}}{c_k+3\sqrt{c_k}})^{-c_k-6\sqrt{c_k}}=\mu(c_k)$
decreases as $c_k$ increases.\footnote{\textcolor{black}{Appendix C gives the proof that $\mu(c_k)$ is a decreasing function.}
}

Similarly, putting $E[X]=c_k+3\sqrt{c_k}$ and $\delta = \frac{3\sqrt{c_k}}{c_k+3\sqrt{c_k}}$ into \eqref{probability_lower},
we can get:
{\small{
\begin{flalign}
&\textbf{Pr}(X < (1-\frac{3\sqrt{c_k}}{c_k+3\sqrt{c_k}})(c_k+3\sqrt{c_k})) = \textbf{Pr}(X < c_k) \nonumber\\
\le &\left[\frac{e^{{-3\sqrt{c_k}}/({c_k+3\sqrt{c_k}})}}
{(1+({-3\sqrt{c_k}})/({c_k+3\sqrt{c_k}}))^{1+({-3\sqrt{c_k}})/({c_k+3\sqrt{c_k}})}}\right]^{c_k+3\sqrt{c_k}}\nonumber\\
= &\textcolor{black}{\frac{e^{-3\sqrt{c_k}}}{(1+\frac{-3\sqrt{c_k}}{c_k+3\sqrt{c_k}})^{c_k}} 
\le \frac{e^{-3\sqrt{1}}}{(1+\frac{-3\sqrt{1}}{1+3\sqrt{1}})^{1}}} = 0.1991 \label{probability_lower2}.
\end{flalign}
}}
The inequality in \eqref{probability_lower2} is due to the fact that
$c_k\ge 1$ and function
$e^{-3\sqrt{c_k}}(1+\frac{-3\sqrt{c_k}}{c_k+3\sqrt{c_k}})^{c_k}=\rho(c_k)$
decreases as $c_k$ increases.\footnote{\textcolor{black}{
Appendix D gives the proof that $\rho(c_k)$ is a decreasing function.}}

Finally, based on \eqref{probability_upper2} and \eqref{probability_lower2},
we can obtain $\textbf{Pr}(X < c_k   \text{ or } X > c_k+6\sqrt{c_k}  ) < 0.5987$,
and the conclusion.
\end{IEEEproof}
\end{theorem}


\begin{theorem}\label{number_of_trials_4_find_random_seeds}
The probability that reader $R$ can find a suitable random seed
within $6$ tests is almost $1$, and the average number of tests is less than $2.5$.
\begin{IEEEproof}
Given sets $\mathcal{A}$ and $\mathcal{B}$, and also a set
$\mathcal{H}$ of functions from $\mathcal{A}$ to $\mathcal{B}$.
That is, $\mathcal{H}=\{h_r:\mathcal{A}\rightarrow \mathcal{B}|r\in \mathcal{R}\}$.
A random hash function $h_r$
can be created by selecting $r$ randomly from $\mathcal{R}$;
and $r$, also known as a random seed, is a unique identifier for a function in $\mathcal{H}$ \cite{hayashi2016more}.
Thus, a random seed $r$ in $\mathcal{R}$
can be considered as equivalent to a hash function in the set $\mathcal{H}$.

Let $\mathcal{H}=\{h_1,h_2,\cdots, h_I\}$ \textcolor{black}{denote} the set of all possible hash functions.
According to Theorem \ref{pick_probability_OPT-C1}, more than $40\%$
of the functions in $\mathcal{H}$ are capable of selecting at least
$c_k$ tags and at most $c_k+6\sqrt{c_k}$ tags for category $P_k$.
So,
the probability that none of the $6$ different functions chosen independently and uniformly at random from $\mathcal{H}$ is suitable is less than $(1-0.4)^6= 0.0467$.
To find a suitable function,
the average number of tested functions is less than $\sum\nolimits_{i=1}^{\infty} 0.4\times 0.6^{i-1} \times i
=2.5$.
\end{IEEEproof}
\end{theorem}

Theorem \ref{number_of_trials_4_find_random_seeds}
is
verified \textcolor{black}{using}
real IDs from the universe $U$
in Section \ref{verify_number_of_trials_4_find_random_seeds}.
Next, we shall show OPT-C1 can pick
any $l$-subset of $P_k$
with the same probability,
where $l\in[c_k,c_k+6\sqrt{c_k}]$,
and then we analyze the communication cost of OPT-C1.

\begin{theorem}\label{pick_probability_subset_OPT-C1}
Let $P_k^{\textbf{s}}$ 
be the subset of $l$ tags selected by round $\mathbb{I}_k$
from $P_k$,
then $P_k^{\textbf{s}}$ has the same probability of $1/(_{\,\,l}^{n_k})$ to be
any $l$-subset of $P_k$.
\begin{IEEEproof}
{{
Every tag has the same probability to be in $P_k^{\textbf{s}}$,
and $|P_k^{\textbf{s}}|=l$,
so $P_k^{\textbf{s}}$ has the same probability of being any one of
the $l$-subsets of $P_k$.
As $P_k$ has $(_{\,\,l}^{n_k})$
$l$-subsets, the probability that $P_k^{\textbf{s}}$
equals to a particular $l$-subset of $P_k$
is ${1}/{(_{\,\,l}^{n_k})}$.
}}
\end{IEEEproof}
\end{theorem}

\begin{theorem}\label{communication_cost_OPT-C1}
Let $|\text{OPT-C1}|$ represent the communication cost of
OPT-C1, then
we have $|\text{OPT-C1}|=\sum\nolimits_{k=1}^K({ \log_2{(96/\log_2n_k)} +\log_2 c_k })$.
\begin{IEEEproof}
In each round $\mathbb{I}_k$, reader $R$
needs to send two parameters: $r$ and $c_k$.
According to Ref. \cite{Mitzenmacher2008Why},
a $(2\log_2(n_k))$-bit random seed $r$ is sufficient to describe a universal hash function.
However, RFID systems can further reduce this cost
because the $96$-bit IDs of tags
contain a high percentage of random bits,
each drawn independently and uniformly at random from
$\{0,1\}$.
Note that the validity of this fact has been verified not only by real experiments with real tag IDs in Section \ref{realize_hash_on_tag} of this paper,
but also by numerous existing RFID protocols that rely on this fact, e.g.,
\cite{liu2018efficient,9086079,liu2022revisiting,zhu2020collisions,9082674,
8948356}.
Therefore, we can pick a $\log_2{(n_k)}$-bit substring of tag $t$'s ID,
and take the value of this substring as the hash value $h(t)$.
This approach costs us at most $\log_2{(96/\log_2n_k)}$ bits,
because we can describe the starting position of a substring with
 $\log_2{(96/\log_2n_k)}$ bits.
\end{IEEEproof}
\end{theorem}

\subsection{The design and analysis of OPT-C2}\label{sec_opt-c2}
OPT-C2 contains $K$ communication rounds: $\mathbb{Q}_1,\mathbb{Q}_2, \cdots, \mathbb{Q}_K$.
{Each round $\mathbb{Q}_k$, $k=1,2,...,K$,
will randomly draw $c_k$ tags from those that entered the selected state during
round $\mathbb{I}_k$ of OPT-C1 and
assign them a unique reporting order from 1 to $c_k$.
}



First, we list some notations used by OPT-C2.
Let $P_k^{\textbf{s}}$ represent the set
consisting of those tags in category $P_k$
that entered the selected state during round $\mathbb{I}_k$ of OPT-C1.
Let $T_k^{\textbf{r}}$ denote the set consisting of tags that enter the ready state during round $\mathbb{Q}_k$.
Besides, we use $cnt$ to denote a local counter
held by each tag $t\in P_k^{\textbf{s}}$
to store its reporting order, and $cnt$ is initialized to 0.




Second, we describe the tags' states when OPT-C2 starts.
After OPT-C1 has selected slightly more than $c_k$
random tags from $P_k$,
we have the following $3$ initial conditions for round $\mathbb{Q}_k$.
\begin{itemize}[leftmargin=0.55cm]
\item[(1)] The number of tags in $P_k^{\textbf{s}}$ is within $[c_k,c_k+\sqrt{c_k}]$,
i.e., $|P_k^{\textbf{s}}|\in[c_k,c_k+6\sqrt{c_k}]$ (by Theorem \ref{pick_probability_OPT-C1});
\item[(2)] $T_k^{\textbf{r}}=\emptyset$ (no tag is in the ready state when OPT-C2 starts);
\item[(3)] Reader $R$ knows which tags in category $P_k$ are in $P_k^{\textbf{s}}$
($R$ can predict the tags' states as it
knows the $96$-bit IDs and category-IDs of all tags).
\end{itemize}
Based on these initial conditions,
the job of round $\mathbb{Q}_k$ is to
randomly pick exactly $c_k$ tags from $P_k^{\textbf{s}}$
and assign them unique reporting orders in $\{1,2,...,c_k\}$.

Third, the detailed steps of round $\mathbb{Q}_k$ are as follows.
\indent\textbf{Step1:} Reader $R$ broadcasts
two parameters $<r,|P_k^{\textbf{s}}|>$ out to all the tags in $P_k^{\textbf{s}}$,
where $r$ is a random seed and $|P_k^{\textbf{s}}|$ is the size of $P_k^{\textbf{s}}$.
\emph{Note that, unlike OPT-C1, OPT-C2 does not require a suitable random seed,
and any random seed will do.
}
\\
\indent\textbf{Step2:} Upon receiving $<r,|P_k^{\textbf{s}}|>$, each tag $t$ in
 set $P_k^{\textbf{s}}$
computes function $h(t)=H(t,r)\bmod |P_k^{\textbf{s}}|$.\\
\indent\textbf{Step3:} Reader $R$
broadcasts a vector $\textbf{F}$ of $|P_k^{\textbf{s}}|$ bits,
where each bit $\textbf{F}[i]$ is set to $1$ if $|T_k^{\textbf{r}}|+\sum\nolimits_{i=0}^{i-1}\textbf{F}[i] < c_k$
and exactly one tag $t$ from $P_k^{\textbf{s}}$ has $h(t)=i$,
otherwise $\textbf{F}[i]$ is set to $0$.\\
\indent\textbf{Step4:} Upon receiving $\textbf{F}$, each tag in $P_k^{\textbf{s}}$
checks $\textbf{F}[h(t)]$. If $\textbf{F}[h(t)]=1$, $t$ sets
$cnt=cnt+\sum\nolimits_{i=0}^{h(t)}\textbf{F}[i]$ and changes its status from the selected state
to ready state;
otherwise, $t$ sets $cnt=cnt+\sum\nolimits_{i=0}^{|P_k^{\textbf{s}}|-1}\textbf{F}[i]$ and stays in the selected state.\\
\indent\textbf{Step5:}
Reader $R$, who knows the IDs of all tags in $P_k^{\textbf{s}}$,
can predict which tags in $P_k^{\textbf{s}}$ will enter the ready state
and move them from $P_k^{\textbf{s}}$ to $T_k^{\textbf{r}}$.\\
\indent\textbf{Step6:}
If $|T_k^{\textbf{r}}|$ is equal to $c_k$, reader $R$
stops round $\mathbb{Q}_k$ and informs all tags in the selected state to enter the unselected state;
otherwise, $R$ repeats steps 1-5.

\begin{figure}[h]
\centering
\includegraphics[width=8cm]{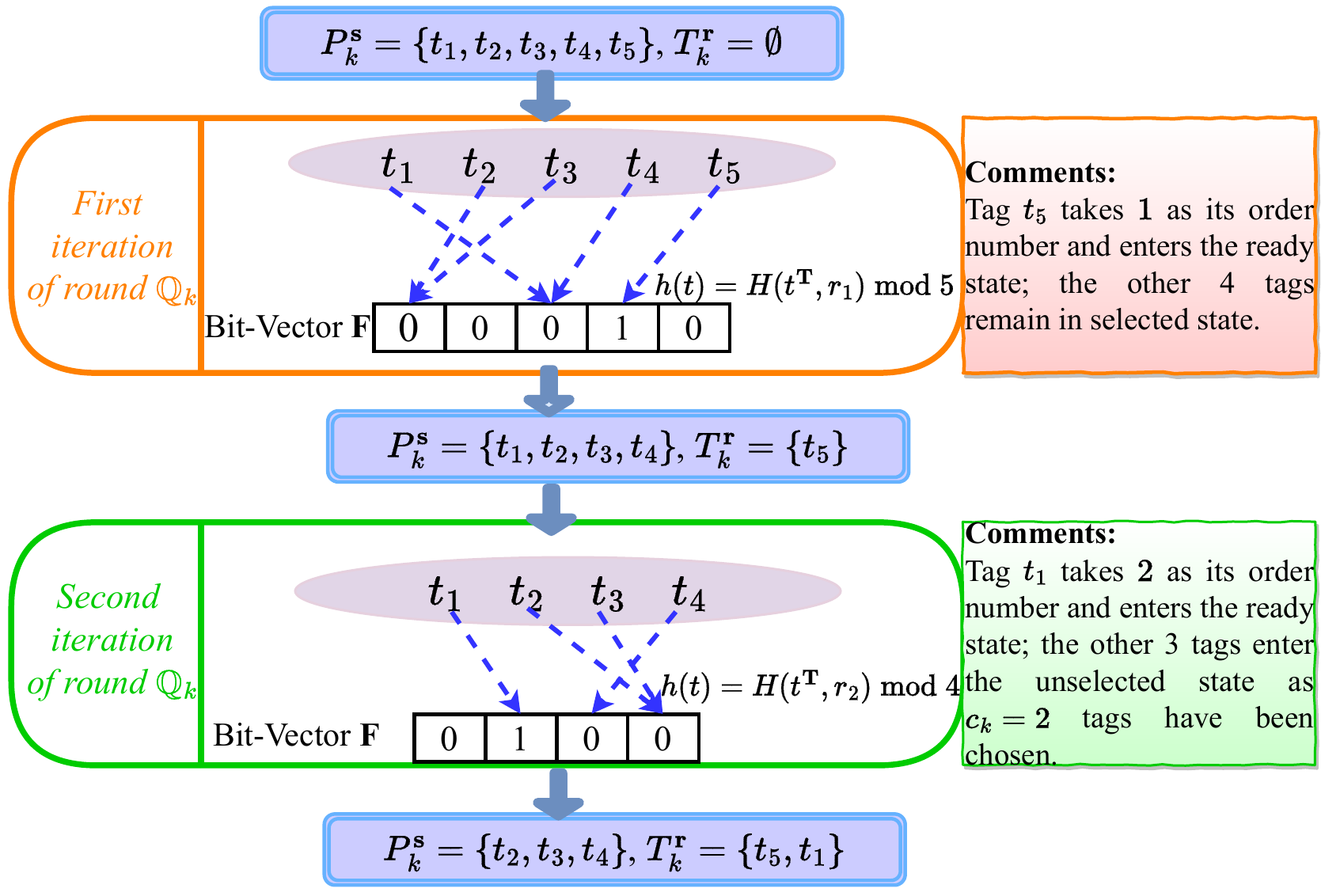}
\caption{A running example of OPT-C2.
}
\label{fig-optc2}
\end{figure}

We take an example to show how OPT-C2 runs.
Suppose, when round $\mathbb{Q}_k$ starts, $5$ tags: $t_1,t_2,\cdots,t_5$ from category $P_k$
stay in the selected state, and $c_k=2$.
Thus, initially, $P_k^{\textbf{s}}=\{t_1,t_2,\cdots,t_5\}$ and
$T_k^{\textbf{r}}=\emptyset$.
Let's look at the first iteration of steps 1-5 in round $\mathbb{Q}_k$.
In step 1, $R$ broadcasts a random seed $r_1$ and $|P_k^{\textbf{s}}|=5$
to all tags in $P_k^{\textbf{s}}=\{t_1,t_2,..,t_5\}$.
Assume that in step 2, the hash values of the $5$ tags are: $h(t_1)=h(t_4)=2$,
$h(t_2)=h(t_3)=0$, and $h(t_5)=3$.
Then, the reader will create a vector $\textbf{F}=[0,0,0,1,0]$ in step 3 and
send it out.
In step 4, since $\textbf{F}[h(t_5)]=1$,
tag $t_5$ takes $1$ as its reporting order and
enters the ready state;
since $\textbf{F}[h(t_1)]=\textbf{F}[h(t_4)]=\textbf{F}[2]=0$,
\textcolor{black}{tags} $t_1$ and $t_4$ update their local counter $cnt$ to $1$ and stay in the selected state;
since $\textbf{F}[h(t_2)]=\textbf{F}[h(t_3)]=\textbf{F}[0]=0$,
\textcolor{black}{tags} $t_2$ and $t_3$ do the same thing as $t_1$ and $t_4$.
In step 5, reader $R$ moves $t_5$ from $P_k^{\textbf{s}}$ to $T_k^{\textbf{r}}$ because tag
$t_5$ is now in the ready state.
Lastly in step 6, since $|T_k^{\textbf{r}}| = |\{t_5\}|=1<c_k=2$,
$R$ repeats step 1-5.
{The second iteration is executed similarly to the first.}
Fig. \ref{fig-optc2} \textcolor{black}{provides a graphic example
of the this process.}


Following, we \textcolor{black}{analyze OPT-C.}

\begin{theorem}\label{picking_ck_OPT-C2}
After round $\mathbb{Q}_k$, exactly $c_k$ tags are in the ready state,
and each of them is assigned a unique reporting order ranging from $1$ to $c_k$.
Moreover, $T_k^{\textbf{r}}$ has the same probability of $1/(_{\,c_k}^{n_k})$ to be
any $c_k$-subset of $P_k$.
\begin{IEEEproof}
First, by step 6 in $\mathbb{Q}_k$, we can observe that reader $R$ stops
when $T_k^{\textbf{r}}$ contains $c_k$ tags from $P^{\textbf{s}}_k$. 
{Also, since each tag always records the number of tags that move from $P^{\textbf{s}}_k$ to $T_k^{\textbf{r}}$
in its local counter $cnt$ (step 4 of OPT-C2), it is easy to determine that the counter $cnt$
on each tag takes a unique integer from $\{1,2,\cdots,c_k\}$.
}
Thus, when  $\mathbb{Q}_k$ ends,
we can conclude:  $|T_k^{\textbf{r}}|=c_k$
and each tag in $T_k^{\textbf{r}}$ is assigned with a unique reporting order.


Second, we analyze the probability that
$T_k^{\textbf{r}}$ equals to
a particular $c_k$-subset of $P_k^{\textbf{s}}$.
{Each tag in $P_k$ has the same probability of being included in $P_k^{\textbf{s}}$,
and each tag in $P_k^{\textbf{s}}$ has the same probability of being included in $T_k^{\textbf{r}}$,
so each tag in $P_k$ has the same probability of being included in $T_k^{\textbf{r}}$.}
Furthermore, since $P_k$ has $(_{\,c_k}^{n_k})$ different
$c_k$-subsets, the probability that $T_k^{\textbf{r}}$
is equal to a particular $c_k$-subset of $P_k$ is $1/(_{\,c_k}^{n_k})$.

\end{IEEEproof}
\end{theorem}

\begin{theorem}\label{communication_cost_OPT-C2}
Let $|\text{OPT-C2}|$ represent the communication cost of OPT-C2, then
we have $|\text{OPT-C2}|\le\sum\nolimits_{k=1}^K{e \times c_k}$.
\begin{IEEEproof}
Suppose $P_k^{\textbf{s}}$ contains $l_k$
tags when round $\mathbb{Q}_k$ starts.
Since OPT-C1 uses a suitable random seed to put tags into $P_k^{\textbf{s}}$
 (Theorem \ref{pick_probability_OPT-C1}), we have $l_k\in[c_k,c_k+6\sqrt{c_k}]$.

First, let's look at the first iteration of $\mathbb{Q}_k$.
In step 1, reader $R$ sends two parameters: $r$ and $|P_k^{\textbf{s}}|$
to create hash function $h()$;
In step 3, $R$ sends a $|P_k^{\textbf{s}}|$-bit vector $\textbf{F}$.
As mentioned \textcolor{black}{earlier} in Theorem \ref{communication_cost_OPT-C1},
a $(2\log_2(|P_k^{\textbf{s}}|))$-bit random seed $r$ is
sufficient to describe a
hash function
\cite{Mitzenmacher2008Why}.\footnote{
A universal hash function mapping to $\{0,1\}^m$ requires
at least $2m$ bits to represents and
can perform nearly as a truly random hash function in practice.}
So, this costs us
$3\log_2(|P_k^{\textbf{s}}|) + |P_k^{\textbf{s}}|\approx |P_k^{\textbf{s}}|$ bits.


Next, we analyze the expected number of tags in $P_k^{\textbf{s}}$
that are moved to set $T_k^{\textbf{r}}$ in
the first iteration.
For any tag $t\in P_k^{\textbf{s}}$, the probability that $t$
occupies the bit $\textbf{F}[h(t)]$ exclusively is $(1-1/|P_k^{\textbf{s}}|)^{|P_k^{\textbf{s}}|-1}\approx e^{-1}$.
Hence, there are on average $e^{-1} |P_k^{\textbf{s}}|$ tags that are moved from
$P_k^{\textbf{s}}$ to $T_k^{\textbf{r}}$,
and then we have
$|P_k^{\textbf{s}}|=(1-e^{-1})l_k$
and $|T_k^{\textbf{r}}|=e^{-1}l_k$ at the end of this iteration.

Following a similar analysis as the first iteration,
we find that $|P_k^{\textbf{s}}|=(1-e^{-1})^J l_k$,
and $|T_k^{\textbf{r}}|=\sum\nolimits_{i=1}^{J} e^{-{\color{black}{1}}}{\color{black}(1-e^{-1})^{i-1}}l_k$ hold true
at the end of the $J$-th iteration, $J\ge2$.
In order to ensure that $|T_k^{\textbf{r}}|=c_k$ (i.e.,
exactly $c_k$ tags enter the ready state),
reader $R$ needs to perform $I$ iterations
to make $|P_k^{\textbf{s}}|=(1-e^{-1})^I l_k$ equal to $ l_k-c_k$.
The total number of bits transmitted in these $I$ iterations is
$\sum\nolimits_{i=0}^{I-1}(1-e^{-1})^i l_k
= e{l_k-e l_k(1-e^{-1})^I}=
e c_k$.
\textcolor{black}{Note that the last equality}
is due to the fact that $I$ satisfies $(1-e^{-1})^I l_k = l_k-c_k$.
Applying this result to the $K$ rounds $\mathbb{Q}_1,\mathbb{Q}_2,...,\mathbb{Q}_K$
completes the proof.
\end{IEEEproof}
\end{theorem}

\setcounter{footnote}{0}

\subsection{The analysis of OPT-C}\label{analysis_OPT-C}
OPT-C first executes OPT-C1 and then OPT-C2.
So according to Theorem \ref{picking_ck_OPT-C2},
OPT-C solves the
\textcolor{black}{CIS problem.}
Its overall communication cost and execution time satisfy the following theorem.
\begin{theorem}\label{th_communication_cost_OPT-C}
Let $|\text{OPT-C}|$ denote the communication cost of OPT-C, then
we have
$|\text{OPT-C}|\le \sum\nolimits_{k=1}^K (\log_2{(\frac{96}{\log_2n_k})} +\log_2 c_k + {e c_k})$. 
Moreover, the execution time of OPT-C
is within
a factor of $\frac{e}{\log_2(e)}\approx 1.88$ of the lower bound ${\rm{T}_{{\rm{lb}}}}$ shown in \eqref{lb_for_perfect_channel_time_lb}.
\begin{IEEEproof}
From Theorem \ref{communication_cost_OPT-C1} and \ref{communication_cost_OPT-C2},
it is clear that $|\text{OPT-C}| = |\text{OPT-C1}| +|\text{OPT-C2}|
\le \sum\nolimits_{k=1}^K (\log_2{(\frac{96}{\log_2n_k})} +\log_2 c_k + {e c_k})$.
Then, the execution time of OPT-C is less than
${\sum\nolimits_{k=1}^K (\log_2{(\frac{96}{\log_2n_k})} +\log_2 c_k + {e c_k})}/{96}\times {\rm{T}}_{{\rm{96}}}$,
where ${\rm{T}}_{{\rm{96}}}$ is the time cost
of transmitting a $96$-bit string from reader $R$ to tags.
Since
{\small{
\begin{flalign}\nonumber
&\frac{{{\sum\nolimits_{k=1}^K \left(\log_2{(\frac{96}{\log_2n_k})} +\log_2 c_k + {e c_k}\right)}}/{96}\times {\rm{T}}_{{\rm{96}}}}
{ {\sum\nolimits_{k=1}^K{\log_2(e)c_k} }/{96}\times {\rm{T}}_{{\rm{96}}}} \nonumber\\
&\approx \frac{{{\sum\nolimits_{k=1}^K \left( {e}\right)c_k}} }
{ {\sum\nolimits_{k=1}^K{\log_2(e)c_k} } }=\frac{e}{\log_2(e)}\approx 1.88,\nonumber
\end{flalign}
}}
we know that
the communication cost of OPT-C
is within a fact of $1.88$ of the lower bound.
Note that the first approximation in the above derivation is due to
the fact that $\frac{\log_2{(96/\log_2 n_k)}}{c_k} < \frac{3.85}{c_k}\ll e$ for $\forall k\in\{1,2,...,K\}$,
and $\log_2(c_k)\ll c_k$.\footnote{In a real RFID system, a category $P_k$ usually contains hundreds of tags, i.e., $n_k\ge 10^2$;
and part (b) of Fig. \ref{fig-reqc} indicates that we need to randomly choose $c_k\in[4,44]$
tags from $P_k$ to ensure a high success probability of $0.99$.}

\end{IEEEproof}
\end{theorem}

\textcolor{black}{Hence, the execution time of OPT-C is mainly determined by $K$ and $c_1,c_2,...,c_K$, it is near-optimal in theory.}


%

\setcounter{footnote}{0}

{\section{\textcolor{black}{The Implementation of OPT-C in Real RFID Systems Using COTS Devices
}}\label{sec_imp}

{\color{black}{To assess the feasibility and performance of the OPT-C protocol, we have implemented it in a real RFID system, utilizing readily available COTS RFID devices. This implementation, denoted as $\text{OPT-C}_\mathbf{IMPL}$, is achieved via reader-end programming without requiring any hardware modifications to the COTS tags. Importantly, $\text{OPT-C}_\mathbf{IMPL}$ exclusively relies
on tags' EPCs and the standard {\tt Select} command,
as mandated by the C1G2 standard \cite{EPCGlobal}.
This ensures seamless integration into real-world RFID systems.\footnote{\textcolor{black}{\textbf{Implementation Rationale:} It is crucial to emphasize that despite OPT-C's utilization of the FSA protocol as the underlying MAC protocol, we deliberately chose to employ the standard {\tt Select} command instead of modifying the FSA protocol. This decision stems from the limited software interface provided by major COTS device manufacturers for reprogramming the FSA protocol to meet our specific requirements \cite{EPCGlobal,ImpSDK}. Moreover, the reprogramming of the {\tt Select} command has demonstrated its efficiency and widespread acceptance within the RFID research community for implementing various other tag management protocols \cite{lin2019tash,an2020acquiring,liu2022time,xie2022efficient}.}}
$\text{OPT-C}_\mathbf{IMPL}$ encompasses two fundamental functionalities, denoted as \textbf{F-1} and \textbf{F-2}, which are outlined as follows:

\indent $\bullet$ \textbf{F-1: Generating Hash Values for Tags without Overburdening COTS Tags:} To achieve this function, we conduct a thorough analysis of the bit composition within the 96-bit IDs (EPCs) of COTS tags. This analysis reveals that the IDs of COTS tags typically contain a sufficient number of random bits. Drawing inspiration from established software implementations of hash functions on COTS tags, as found in references \cite{lin2019tash,an2020acquiring,hu2019bringing,yang2017analog}, we successfully generate the required hash value on each COTS tag. Importantly, this is achieved without imposing any additional computational burden on the COTS tags. Further details are in Section \ref{realize_hash_on_tag}.

$\bullet$ \textbf{F-2: Designing an Efficient Algorithm to Generate Appropriate {\tt "Select"} Commands for Swiftly Selecting Tags with Hash Values Below a Given Threshold $\bm{\tau}$:}
The straightforward approach to selecting all tags with a hash value no more than $\bm{\tau}$ would involve issuing $\bm{\tau}+1$ individual {\tt Select} commands, each designated by an integer $i$ ranging from $0$ to $\bm{\tau}$. However, this approach proves time-consuming, requiring $\bm{\tau}$ individual {\tt "Select"} commands. Therefore, we harness the tag filtering function provided by the {\tt "Select"} command \cite{EPCGlobal,lin2019tash,an2020acquiring} to design a more efficient method that can generate a significantly smaller number of {\tt Select} commands for choosing all tags with hash values no more than $\bm{\tau}$. We specifically design a method called \textbf{SelGen}, which can generate no more than
$\lceil \log_2(\bm{\tau})\rceil+1$ {\tt Select} commands for selecting all tags with hash values no more than $\bm{\tau}$ from each category $P_k$.
Further details are in Section \ref{sec_for_SelectGenerate}.

$\bullet$ With these core functionalities, we effectively implement the proposed OPT-C protocol in a real RFID system consisting of COTS devices. The details are in Section \ref{implementation of opt-c}.

\begin{figure}[!t]
\centering
\includegraphics[height=3.7cm]{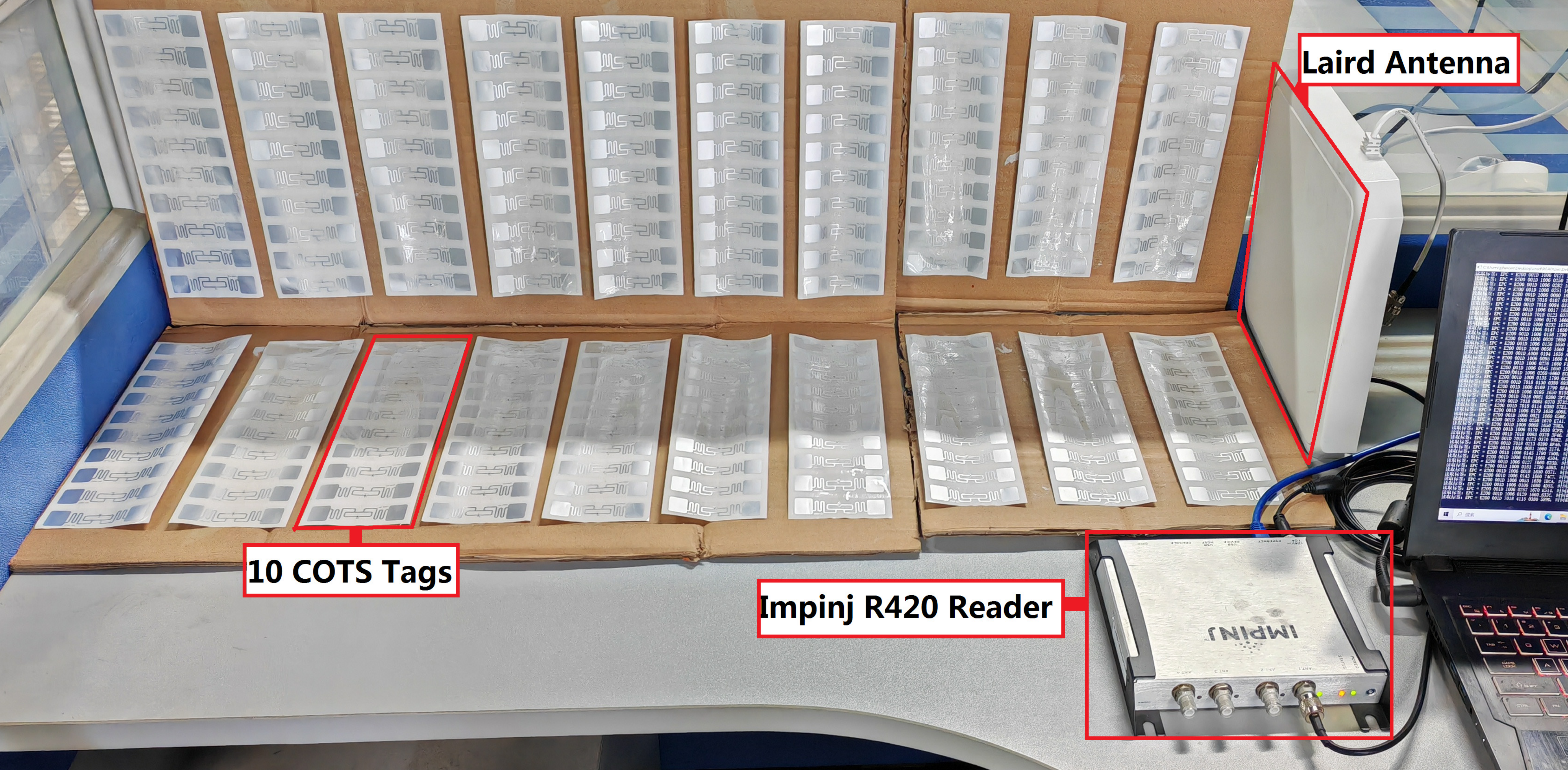}
\caption{\textcolor{black}{Real-world experimental scenario with densely deployed COTS tags in a small area.}}
%
\label{fig_real_scenario_review_appendix}
\end{figure}

\subsection{Generating Hash Values on COTS Tags}\label{realize_hash_on_tag}

This section, we present our approach to generating hash values for COTS tags, a key aspect of the OPT-C protocol. We have observed that the 96-bit IDs (EPCs) of COTS tags typically contain a substantial number of random bits. This observation serves as the foundation for achieving functionality \textbf{F-1} without requiring any hardware modifications to the COTS tags.

To validate this insight, we conduct practical experiments using an Impinj Speedway R420 reader in conjunction with a diverse set of $200$ COTS tags. This set includes a variety of widely used COTS tags:
$100$ ALN-9630 tags, $50$ NXP-Ucode7 tags, and $50$ Impinj-ER62 tags.
Fig. \ref{fig_real_scenario_review_appendix} shows the real RFID system.
After collecting the IDs of these tags, we pursue two critical tasks:

\textbf{1. Quantification of Entropy for IDs (EPCs):} To perform a systematic analysis, we partitioned each tag's EPC into $24$ subarrays, each comprising $4$ bits. Subsequently, we quantified the empirical entropy of these subarrays, shown in Fig. \ref{fig_real_tag_entropy}.
Notably, the $21$st, $22$nd, $23$rd, and $24$th subarrays exhibit entropies of $3.97$,
$3.98$, $3.95$, and $3.94$ bits, respectively. This indicates the presence of more than $3.9$ independently and uniformly distributed random bits in each subarray.

\textbf{2. Generation of Hash Values from Random Bits:} We delineate the approach for generating hash values using these identified random bits. Let $P_k$ represent a subset of tags ($k\in\{1,2,\cdots,K\}$),
encompassing any $n_k$ of the 200 COTS tags, and let $\mathcal{L}=\lceil(\log_2(n_k)\rceil)$. We expound on how to generate a hash value ($h(t)$) for each tag $t\in P_k$ using its EPC (composed of $96$ bits denoted as
$A[1], A[2], \cdots, A[96]$). It's noteworthy that the subarrays spanning from the $21$st to the $24$th
position ($A[81], A[82], \cdots, A[96]$) are recognized to contain approximately $16$ random bits,
which are deemed $16$ independent and uniformly distributed random bits.
Consequently, to create the hash value $h(t)$ for tag $t$, we execute the following two steps:\\
\textbf{Step 1}: Randomly select an integer $\mathcal{I}$ from the range ${81, 82, \ldots, 96-\mathcal{L}+1}$.\\
\textbf{Step 2}: Employ the decimal value of the bit array: $A[\mathcal{I}], A[\mathcal{I}+1], \cdots, A[\mathcal{I}+\mathcal{L}-1]$, as the hash value $h(t)$ specific to the tag $t$ in question.

This approach ensures the generation of hash values without overburdening the COTS tags, aligning with the requirements of the OPT-C protocol.\footnote{\textcolor{black}{
It's crucial to note that the number of random bits, $\mathcal{L}$, utilized for generating the hash value depends on the value of $n_k$ (representing the number of tags within category $P_k$).
Since the number of tags within the reader's interrogation range typically remains below $10^5$ \cite{chu2021efficient,9086079,liu2022revisiting}.
Consequently, the value of $n_k$ is significantly smaller than $10^5$.
Thus, the utilization of these $16$ random bits effectively suffices for generating the required hash values, aligning with the OPT-C protocol's stipulated requirements.}}
}}

{\color{black}{\subsection{Efficient Generation of {\tt Select} Commands for Hash-Based Tag Selection}\label{sec_for_SelectGenerate}

Building upon the generated hash values in Section \ref{realize_hash_on_tag},
we propose an approach for the efficient generation of {\tt Select} commands,
designed to select tags in a category $P_k$ whose hash values are
less than or equal to a specified threshold $\bm{\tau}$. This approach fulfills functionality \textbf{F-2}.

\begin{figure}[!t]
\centering
\includegraphics[height=3.7cm]{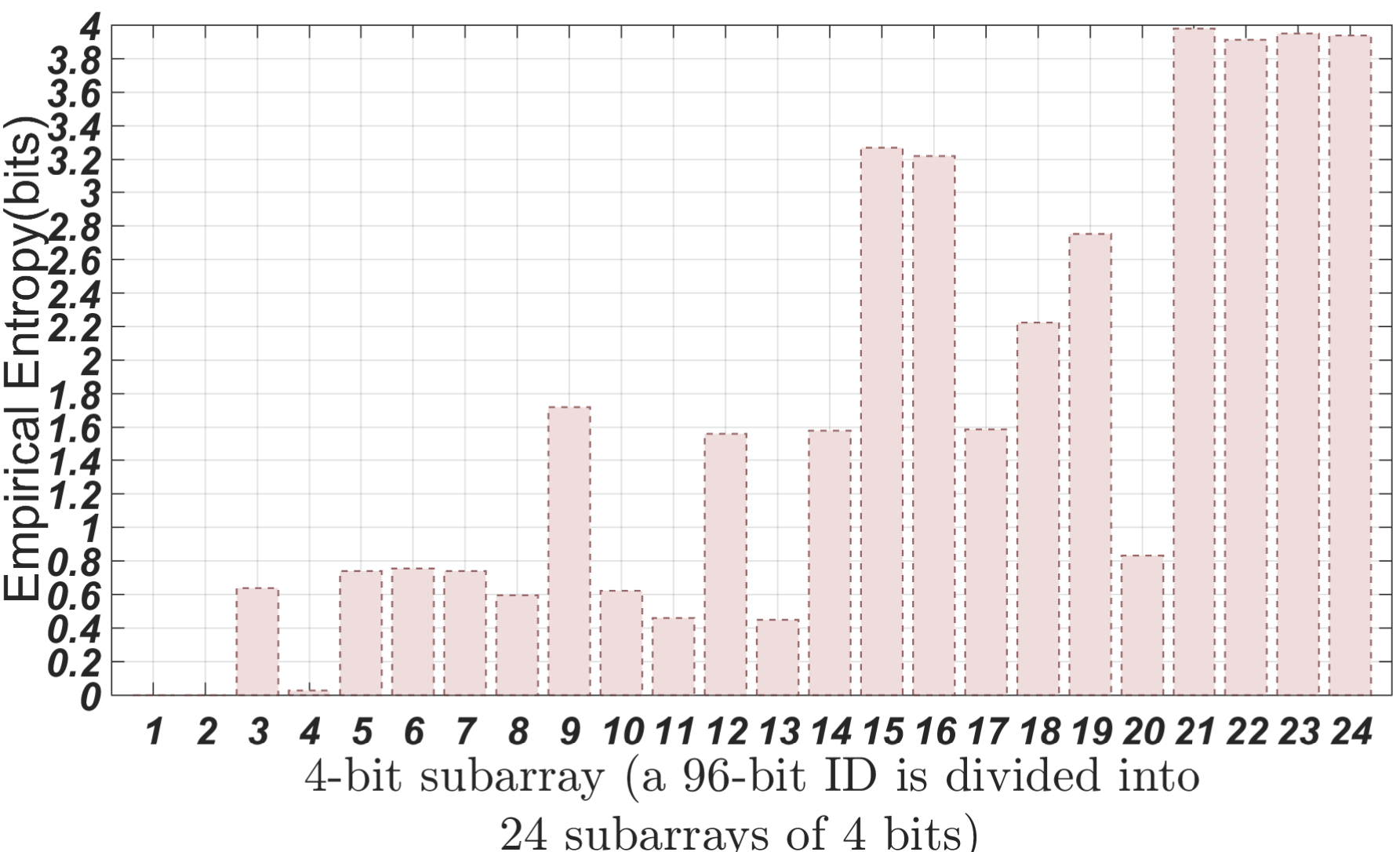}
\caption{\textcolor{black}{The entropy of $200$ COTS tag IDs.}}
\label{fig_real_tag_entropy}
\end{figure}

As defined by the C1G2 standard \cite{EPCGlobal}, a COTS RFID reader ($R$) can employ a {\tt Select} command to specify a particular group of tags for an upcoming tag inventory process. The {\tt Select} command defines a filter string against which tags are compared to determine if they meet the criteria, and only the matching tags are included in the upcoming inventory. Consequently, for a given threshold value $\bm{\tau}$, reader $R$ can execute $\bm{\tau}+1$ {\tt Select} commands, each using a filter string $i$ to select tags with hash values equal to $i$, where $i\in\{0,1,\cdots,\bm{\tau}\}$. However, due to the common prefixes shared by hash values within $\{0,1,\cdots, \bm{\tau}\}$ in binary form, the number of {\tt Select} commands can be significantly reduced. For instance, when using a $4$-bit array to represent the hash value of each tag, a single {\tt Select} command with the filter string $\mlq\mlq000*\mrq\mrq$ can select tags with hash values not exceeding 1. This is because the tag with a hash value of $0$ ($\text{0000}_2$) and the tag with a hash value of 1 ($\text{0001}_2$) share the same unique prefix $\mlq\mlq 000\ast\mrq\mrq$.

To address this optimization, we introduce a method, called \textbf{SelGen}, which
generates no more than $\lceil\log_2(\bm{\tau})\rceil+1$ {\tt Select} commands to choose all tags
with hash values not exceeding $\bm{\tau}$.
Specifically, assume that each tag uses a $\mathcal{L}$-bit array in
its EPC memory as its hash value\footnote{\textcolor{black}{Each tag uses the bit
array
$A[\mathcal{I}], A[\mathcal{I}+1], \cdots, A[\mathcal{I}+\mathcal{L}-1]$
in its EPC to generate its hash value.}},
and $(\tau_{\mathcal{L}}, \tau_{\mathcal{L}-1}, \cdots, \tau_1)$ denotes the binary form of $\bm{\tau}$ ($\bm{\tau}=\sum\nolimits_{i=1}^{\mathcal{L}} \tau_i 2^{i-1} $), the \textbf{SelGen} method proceeds as below:\\
\emph{
\noindent{(1)} Reader $R$ initializes an empty set $\mathbb{S}$.\\
\noindent{(2)} For each $i\in\{\mathcal{L},\mathcal{L}-1,\cdots,1\}$\\
{(3)}$\quad$If $\tau_i=1$, $R$ generates a {\tt Select} command using $\mlq\mlq \tau_{\mathcal{L}} \tau_{\mathcal{L}-1} \cdots \tau_{i+1}
\underbrace{0}_{1} \underbrace{\ast\ast\cdots\ast}_{i-1}\mrq\mrq$ as the filter string and adds this command to $\mathbb{S}$.\\
{(4)}$\quad$If
$\tau_{i-1}=\tau_{i-2}=\cdots=\tau_1=1$ or $i=1$, $R$
generates a {\tt Select} command using $$\mlq\mlq {\tau_{\mathcal{L}} \tau_{\mathcal{L}-1} \cdots \tau_{i+1}} \underbrace{\tau_i}_{1} \underbrace{\ast\ast\cdots\ast}_{i-1}\mrq\mrq$$ as the filter string, adds this command to $\mathbb{S}$, and ends the loop.
}

The filter strings generated in step (3) of \textbf{SelGen} can select
any hash value less than $\bm{\tau}$.
This is because, for any value
$\bm{x} = (x_{\mathcal{L}}, x_{\mathcal{L}-1}, \cdots, x_1) < \bm{\tau} = (\tau_{\mathcal{L}}, \tau_{\mathcal{L}-1}, \cdots, \tau_1)$, there exists an index $j\in\{ \mathcal{L}, \mathcal{L}-1, \cdots, 1 \}$ such that $\tau_{\mathcal{L}} = x_{\mathcal{L}}, \tau_{\mathcal{L}-1} = x_{\mathcal{L}-1}, \cdots, \tau_{j+1} = x_{j+1}$, and $\tau_{j} > x_{j}$.
Since step (4) shall select the value $\bm{\tau}$, we know that \textbf{SelGen}
can ensure the selection of all tags with hash values not exceeding $\bm{\tau}$.
Moreover, \textbf{SelGen} generates at most a total of $\sum\nolimits_{i=1}^{\mathcal{L}} {\tau_i} + 1$ {\tt Select} commands, corresponding to the number of $\mlq 1\mrq $s in the string $(\tau_{\mathcal{L}}, \tau_{\mathcal{L}-1}, \cdots, \tau_1)$ plus one. Therefore, the number of {\tt Select} commands in $\mathbb{S}$ is no more than $\lceil\log_2(\bm{\tau})\rceil+1$.\footnote{\textcolor{black}{We would like to provide details on the structure of the
{\tt Select} commands generated by \textbf{SelGen}. As per the C1G2 standard \cite{EPCGlobal},
a {\tt Select} command utilizes four fields:
\underline{MemBank},
\underline{Pointer},
\underline{Length}, and
\underline{Mask}
to define the filter string (see \textbf{Appendix E}).
Thus, when each tag in a category uses
the $A[\mathcal{I}], A[\mathcal{I}+1],
\ldots, A[\mathcal{I}+\mathcal{L}-1]$ in its
EPC memory bank as its hash value, step (3) of \textbf{SelGen} adds:
${\texttt{Select}}(\text{\underline{MemBank}}=1,
\text{\underline{Pointer}}=\mathcal{I}, \text{\underline{Length}}=\mathcal{L}-i+1,
\text{\underline{Mask}}=\mlq\mlq \tau_{\mathcal{L}} \tau_{\mathcal{L}-1} \cdots \tau_{i+1}0\mrq\mrq)$
to $\mathbb{S}$. Similarly,
step (4) of \textbf{SelGen} adds:
${\texttt{Select}}(\text{\underline{MemBank}}=1,
\text{\underline{Pointer}}=\mathcal{I}, \text{\underline{Length}}=\mathcal{L}-i+1,
\text{\underline{Mask}}=\mlq\mlq \tau_{\mathcal{L}} \tau_{\mathcal{L}-i+1} \cdots \tau_{i+1}\tau_{i}\mrq\mrq)$
to $\mathbb{S}$.
Note that step (4) is used
to ends the loop when
all the unprocessed bits are $1$s
or when all bits has been processed.
}
}

\subsection{Implementation of OPT-C} \label{implementation of opt-c}

In this section, we detail the implementation of the OPT-C protocol,
denoted as $\text{OPT-C}_\mathbf{IMPL}$,
using the hash value generation method discussed
in Section \ref{realize_hash_on_tag} and the \texttt{Select}
command generation method (\textbf{SelGen}) explained in
Section \ref{sec_for_SelectGenerate}.

Let $S$ represent the tag population, which comprises 200 COTS tags categorized into $K$ distinct groups: $P_1, P_2, \cdots, P_K$. Each category $P_k$ contains $n_k$ tags selected from the population.
The $\text{OPT-C}_\mathbf{IMPL}$ implementation is built based on
the functionalities \textbf{F-1} and \textbf{F-2}.
Specifically, we first employ the last $16$ bits ($A[81], A[82], \ldots, A[96]$)
in each tag's EPC to generate the hash value, denoted as $h(t)$, for every tag $t$ in a category $P_k$, where $k \in \{1, 2, \ldots, K\}$.\footnote{\textcolor{black}{In scenarios where tags may occasionally lack a sufficient number of random bits, a practical solution is available. Additional random bits can be conveniently pre-written into the User memory of the tag to ensure the creation of the necessary hash value. This approach has been adopted by several leading works in the RFID systems domain, as evidenced by references \cite{8718354,an2020acquiring,liu2023batch,9771437}.}}
Second, we employ the \textbf{SelGen} method to
implement the two main stages of the OPT-C protocol in a real RFID system as follows.

$\bullet$ \emph{Implementing the first stage of OPT-C:}\\
The initial stage, OPT-C1, encompasses a coarse sampling process that
 draws roughly $c_k$ tags from each category $P_k$ (where $k \in \{1, 2, \cdots, K\}$).
  The primary objective is to select all tags with a hash value that does not exceed the defined threshold $\bm{\tau} = c_k + 3\sqrt{c_k}$. To accomplish this, the reader $R$ generates a hash value for each tag $t$ in $P_k$ by extracting $\lceil\log_2(n_k)\rceil$ bits from the tag's EPC. Subsequently, the \textbf{SelGen} method is invoked with a threshold set at $\bm{\tau} = c_k + 3\sqrt{c_k}$ to create a set of \texttt{Select} commands, designated as $\mathbb{S}_k^{\textbf{\Rmnum{1}}}$, capable of selecting all tags with hash values not exceeding $\bm{\tau}$. $R$ then executes the commands in $\mathbb{S}_k^{\textbf{\Rmnum{1}}}$ to select a set, $P_k^{\textbf{s}}$, comprising slightly more than $c_k$ tags from $P_k$ (refer to Section \ref{opt-c1} for further details).


$\bullet$\emph{Implementing the second stage of OPT-C:}\\
The second stage, OPT-C2, involves a more refined sampling process, which precisely selects $c_k$ tags from $P_k^{\textbf{s}}$ and assigns them unique reporting orders. To implement OPT-C2, the reader $R$ generates a hash value for each tag $t$ within $P_k^{\textbf{s}}$ by extracting a hash value containing
$\log_2(|P_k^{\textbf{s}}|^2) \leq 2\log_2(c_k+6\sqrt{c_k})$ random bits from the tag's EPC.\footnote{\textcolor{black}{The rationale for generating a $\log_2(|P_k^{\textbf{s}}|^2)$-bit hash value for each tag, ranging from 0 to $|P_k^{\textbf{s}}|^2-1$, is as follows.
When $\mathcal{N}$ objects are hashed to integers
in $\{0,1,\cdots,\mathcal{N}^2-1\}$, the probability of at least one pair of objects sharing the same hash value is no more than $(^{\mathcal{N}}_{2})\frac{1}{{\mathcal{N}}^2}\le 0.5$. Consequently, the probability of no two objects sharing the same hash value exceeds $0.5$.
Taking this into consideration, by adopting the hash value range $\{0,1,\cdots,|P_k^{\textbf{s}}|^2-1\}$, we can easily identify an index $I$ from the set $\{81,82,\dots, 96-\mathcal{L}+1\}$ that guarantees that exactly
$c_k$ tags within the set $P_k^{\textbf{s}}$ have hash values not exceeding $\bm{\tau}$.
Here, $\bm{\tau}$ is designated as the $c_k$-th smallest hash value among all the tags in $P_k^{\textbf{s}}$.
}}
Subsequently, $R$ leverages the \textbf{SelGen} method to produce a set of \texttt{Select} commands, denoted as $\mathbb{S}_k^{\textbf{\Rmnum{2}}}$, which selectively pick precisely $c_k$ tags from $P_k^{\textbf{s}}$.
Each chosen tag is implicitly assigned a unique reporting order in the form of a unique integer.
Finally, $R$ executes the commands within $\mathbb{S}_k^{\textbf{\Rmnum{2}}}$ to select exactly $c_k$ tags from $P_k^{\textbf{s}}$.
Please note that these $c_k$ tags constitute the set $T_k^{\textbf{r}}$ (see Section \ref{sec_opt-c2}).
}}

\vspace{-0.3cm}
\section{\textcolor{black}{Simulations}}\label{simulation_results}

In this section,
\textcolor{black}{we evaluate OPT-C's performance}
with two \textcolor{black}{state-of-the-art} category information sampling protocols: TPS \cite{liu2018efficient} and
ACS \cite{chu2021efficient}.
We also verify Theorem \ref{number_of_trials_4_find_random_seeds}
\textcolor{black}{using} $96$-bit IDs from the universe $U$ to show
that reader $R$ can find a suitable random seed rapidly in practice.

All simulations are conducted based on
the C1G2 standard \cite{EPCGlobal} to measure communication times for the tested protocols. Our setup includes a transmission rate of $26.7$ kbps from reader $R$ to the tags, with a $302\mu$s interval between consecutive transmissions. Therefore, reader $R$ requires $\rm{T}_{\rm{96}}=
3897.5\times10^{-6}$s to transmit a 96-bit string to the tags. Subsequently, the time needed to send an $f$-bit string to the tags was $\frac{f\rm{T}_{\rm{96}}}{96}$s.\footnote{\textcolor{black}{Please note that, the time required for the reader $R$
to access the selected tags for their category information is not included in the protocol's execution time,
as this cost remained consistent across all protocols.
Let $\rm{T_{CI}}$ denote the time required for a tag to transmit its category information.
The total time for collecting information from all categories is $\sum\nolimits_{k=1}^K {\rm{T_{CI}}}\times c_k$,
which remains constant regardless of which protocol is used.
}}

\textcolor{black}{We refrain from evaluating the success probability, specifically the likelihood
that at least one of the $c_k$ sampled tags from $P_k$ is not missing.
This is because all three protocols exhibit the same success probability while
demonstrating distinct execution times with identical reliability numbers.
However, in cases where an equal running time budget is assigned to all
tested protocols, OPT-C surpasses others in success probability. This is
attributed to its capability to sample a greater number of tags within
the same time frame,
as depicted in Figs.
\ref{fig_for_scenK},
\ref{fig_for_scenN},
\ref{fig_for_scenC},
and \ref{fig_fpr}.
}

\subsection{\textcolor{black}{Simulation results} for execution times}
Fig. \ref{fig_for_scenK} shows the results of the three protocols in Scenarios 1 and 2.
Please note that the lower bound $\rm{T}_{\rm{lb}}$ in the two figures
is drawn according to \eqref{lb_for_perfect_channel_time_lb} in Theorem \ref{lowerbound}.
From this figure, we can observe:
(1) the execution time of OPT-C increases with $K$, which is
consistent with its theoretical time shown in Theorem \ref{th_communication_cost_OPT-C};
and (2) the execution time of OPT-C is significantly
less than the other protocols, and is actually close to the lower bound of execution time.
For example, in Scenario 1, when $K=100$ (the number of categories is $100$)
the execution times of OPT-C, ACS, and TPS are  $1.5$s, $3.3$s and $6.1$s,
respectively;
in Scenario 2, when $K=100$,
the execution times of OPT-C, ACS, and TPS are  $0.6$s, $1.2$s and $2.2$s,
respectively.
Generally, in Scenarios 1 and 2,
OPT-C reduces the execution time by a factor of about $2.1$,
and
is within about $1.88$ times of the lower bound $\rm{T}_{\rm{lb}}$.

\begin{figure}[!t]
\centering
\includegraphics[height=4.5cm]{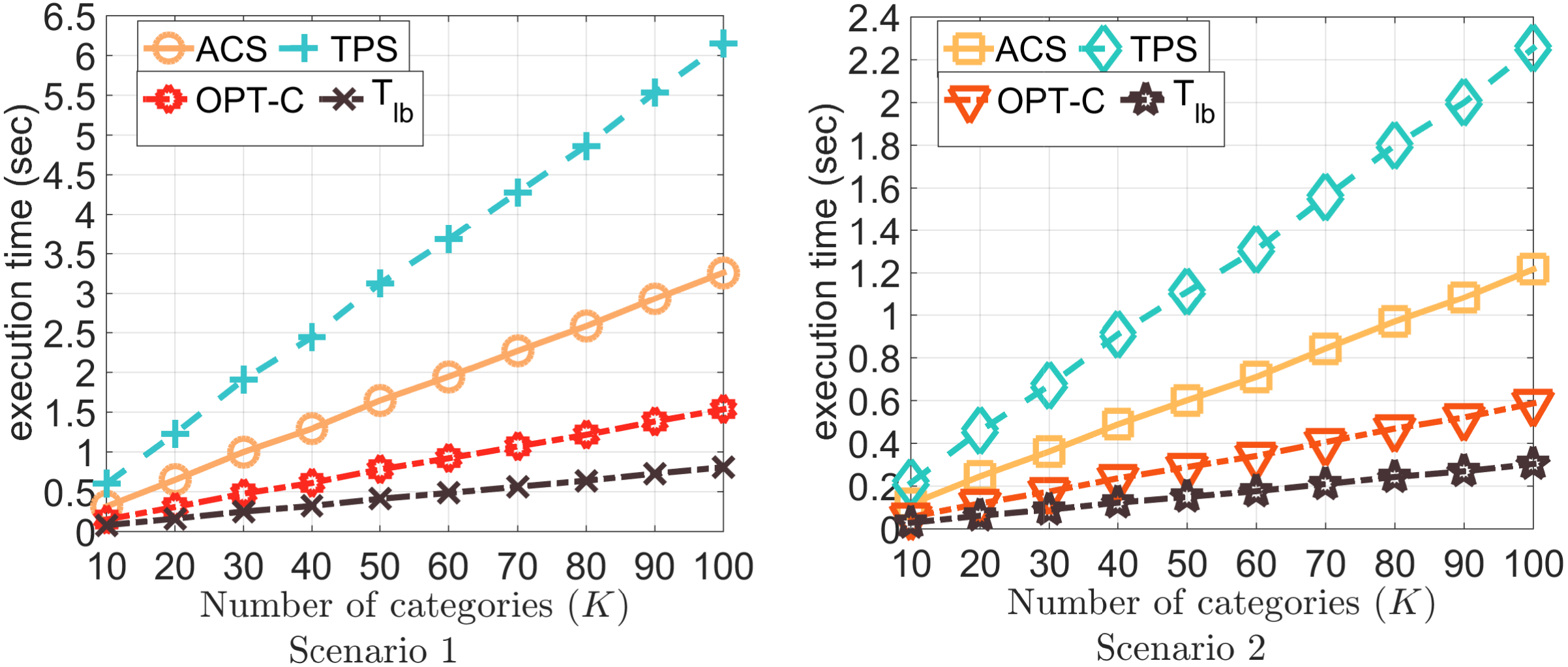}
\caption{{Performance comparisons for Scenarios 1 and 2 with varied $K$ (the number of categories).}}
\label{fig_for_scenK}
\end{figure}

\begin{figure}[!t]
\centering
\includegraphics[height=4.5cm]{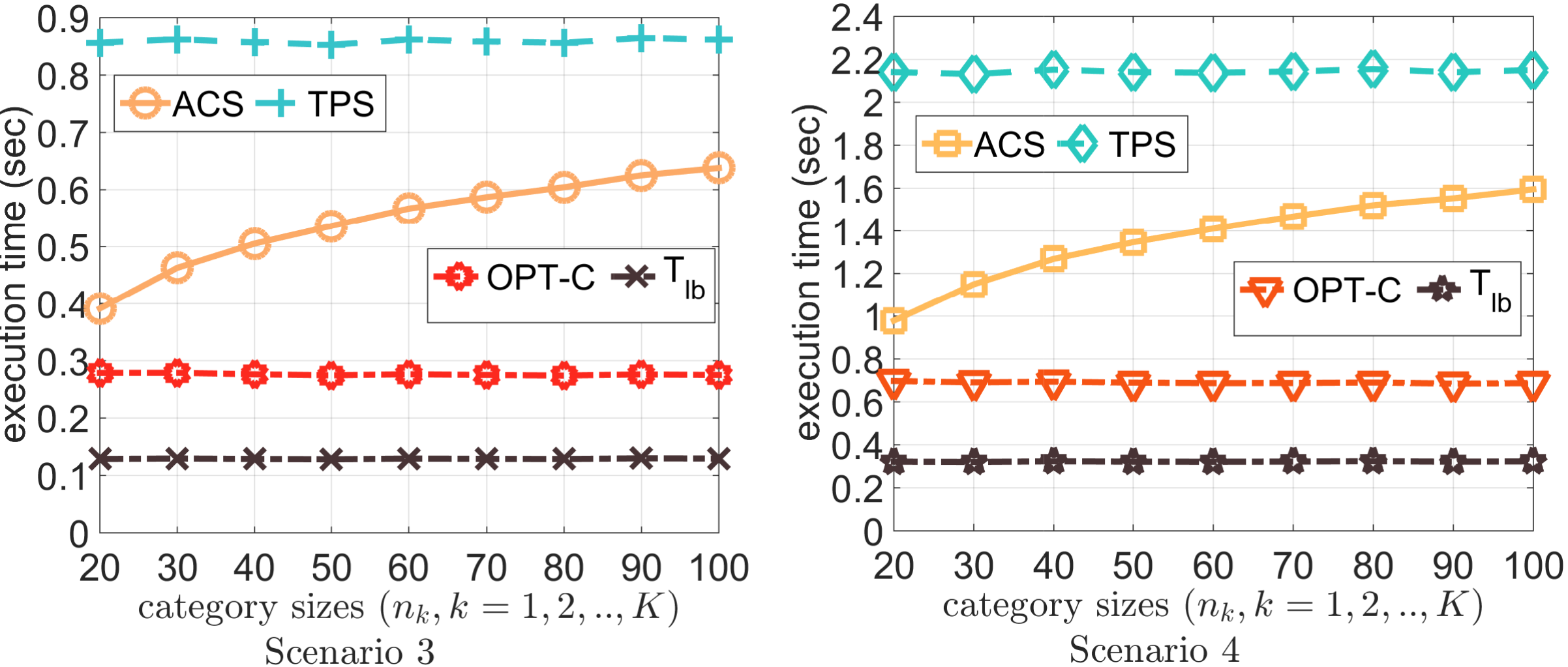}
\caption{{Performance comparisons for Scenarios 3 and 4 with varied $n_1,n_2,...,n_K$
(category sizes).}}
\label{fig_for_scenN}
\end{figure}

\begin{figure}[!t]
\centering
\includegraphics[height=4.5cm]{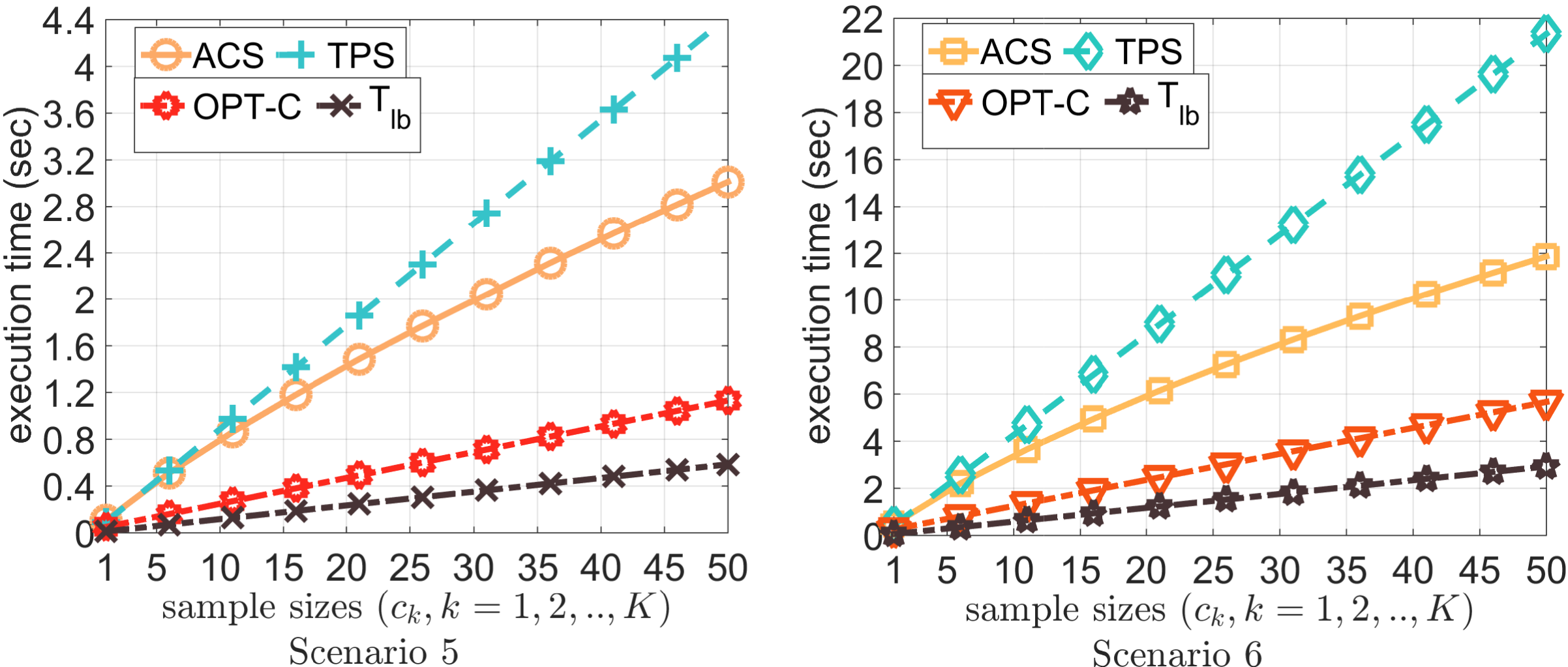}
\caption{{Performance comparisons for Scenarios 5 and 6 with varied sample size $c_k,k=1,2,...,K$.}}
\label{fig_for_scenC}
\end{figure}


Next, for Scenarios 3 and 4,
we show the results of the three protocols in Fig. \ref{fig_for_scenN}.
This figure tells: (1) the execution time of OPT-C remains unchanged with the increase of the $K$ category sizes, which is consistent
with its theoretical time shown in Theorem \ref{th_communication_cost_OPT-C};
and (2) the execution time of OPT-C is much
less than other protocols, and close to the lower bound.
For example, in Scenario 3, when $n_1=n_2=\cdots=n_K=100$,
the execution times of OPT-C, ACS, and TPS are  $0.3$s, $0.6$s and $0.8$s,
respectively;
in Scenario 4, when $n_1=n_2=\cdots=n_K=100$,
the execution times of OPT-C, ACS, and TPS are  $0.7$s, $1.6$s and $2.2$s,
respectively.
Generally, in Scenarios 3 and 4, OPT-C reduces the execution time by a factor of about $2.2$, and
is within about $1.88$ times of the lower bound  $\rm{T}_{\rm{lb}}$.

Lastly, for Scenarios 5 and 6,
we show the results of the three protocols in Fig. \ref{fig_for_scenC}.
This figure demonstrates: (1) the execution time of OPT-C increases with the $K$ sample sizes,
which is also indicated by its theoretical time shown in Theorem \ref{th_communication_cost_OPT-C};
and (2) the execution time of OPT-C is much less than
other protocols, and is close to the lower bound.
For example, in Scenario 5, when $c_1=c_2=\cdots=c_K=50$,
the execution times of OPT-C, ACS, and TPS are  $1.1$s, $3$s and $4.4$s,
respectively;
in Scenario 6, when $n_1=n_2=\cdots=n_K=100$,
the execution times of OPT-C, ACS, and TPS are  $5.7$s, $11.9$s and $21.3$s,
respectively.
Overall for Scenarios 5 and 6, OPT-C reduces the communication time by a factor of about $2.6$,
and is within about $1.88$ times of
the lower bound $\rm{T}_{\rm{lb}}$.



The experimental results of scenarios 1-6 show that OPT-C significantly
outperforms the other two protocols in terms of execution time and can approach the lower bound $\rm{T}_{\rm{lb}}$.\footnote{\textcolor{black}{The OPT-C protocol demonstrates favorable scalability within for RFID systems with a large number of tags, as its execution time only
grows linearly with
the $K$ sample sizes which are usually not excessively large, as indicated by Fig. \ref{fig-reqc}.}
}
This verifies the theoretical result in Theorem \ref{th_communication_cost_OPT-C}.
There are two reasons why OPT-C is superior: its novel coarse sampling protocol,
which can select a little more than $c_k$ tags from each category $P_k$
at an almost negligible time cost (see Theorem \ref{communication_cost_OPT-C1}),
and its refined sampling protocol,
which can extract exactly $c_k$ tags from these already selected tags
with a time cost that is near the lower bound
(Theorem \ref{communication_cost_OPT-C2}).

}
%
%
%
%
%
%

\begin{figure}[!t]
\centering
\includegraphics[height=4.5cm]{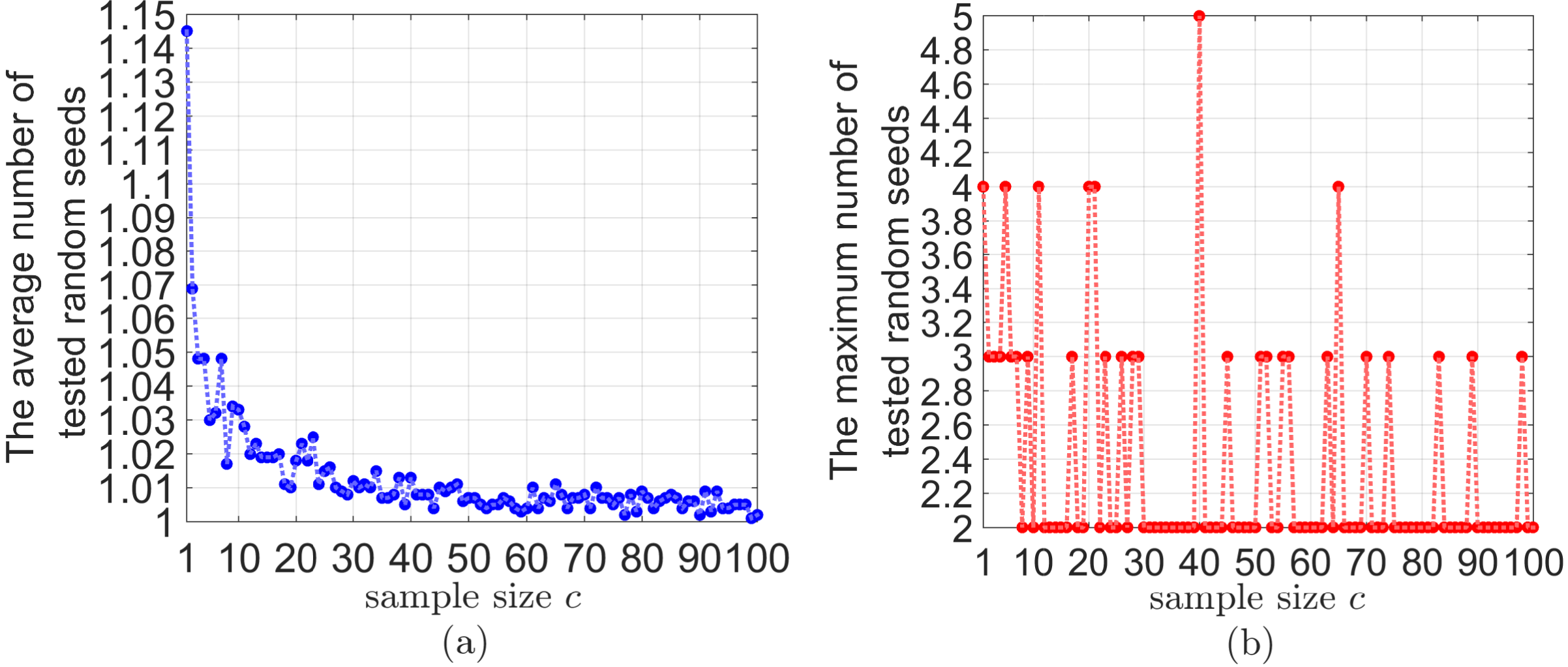}
\caption{{The number of tested random seeds in Scenario 7 ($n=10^3$, $c\in\{1,2,\cdots,100\}$)
for the basic sampling problem.\protect\footnotemark[9]
(a) Average Number. (b) Maximum Number.
}}
\label{fig_for_sc5_experiment}
\end{figure}

\footnotetext[9]{\label{fig_note}\textcolor{black}{In the basic sampling problem, there is only one category $P$ of $n$ tags from $U$, and the objective is to select at least $c$ and at most $c+6\sqrt{c}$ tags randomly from $P$.}}

\begin{figure}[!t]
\centering
\includegraphics[height=4.5cm]{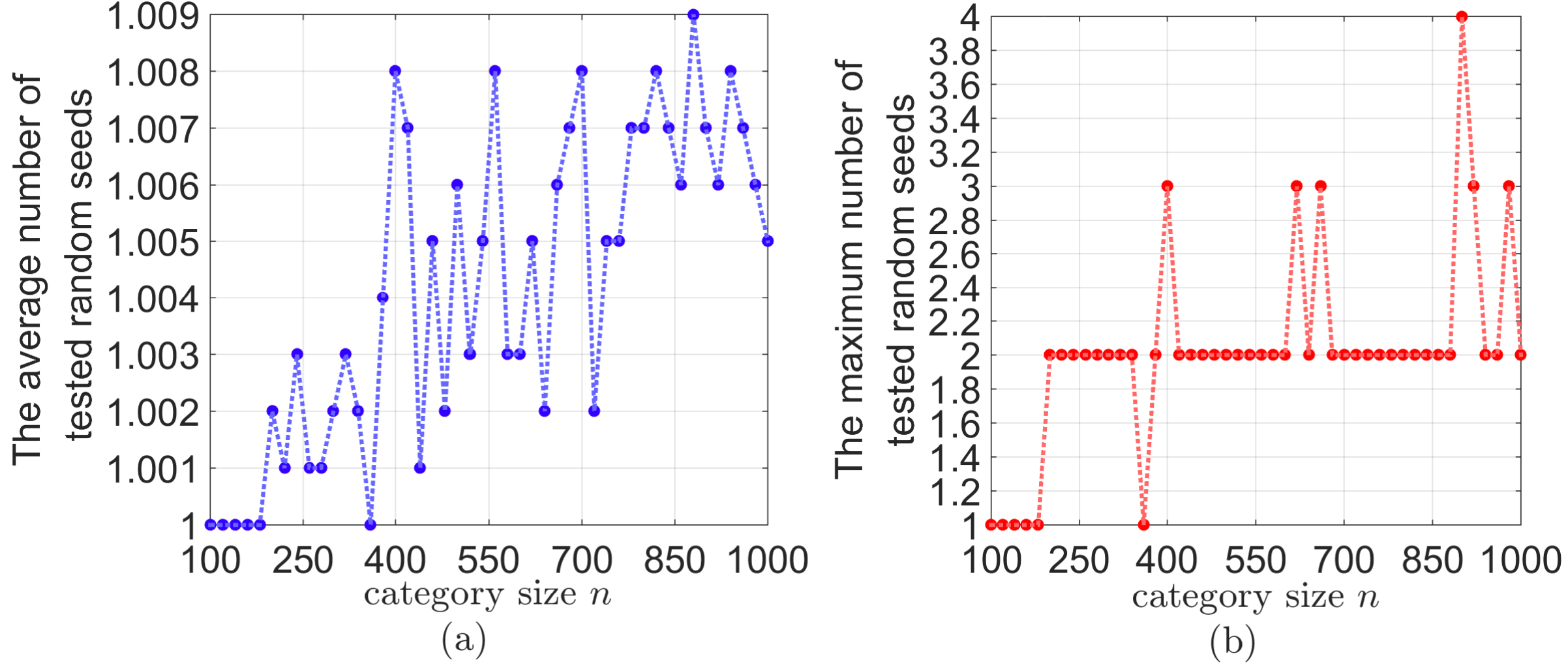}
\caption{{The number of tested random seeds for Scenario 8 ($c=50$, $n\in\{100,120,\cdots,10^3\}$)
for the basic sampling problem.\protect\footnotemark[9]
(a) Average Number. (b) Maximum Number.
}}
\label{fig_for_sc6_experiment}
\end{figure}

\setcounter{footnote}{0}
{\color{black}{
\subsection{Verifying Theorem \ref{number_of_trials_4_find_random_seeds}} \label{verify_number_of_trials_4_find_random_seeds}
We investigate
the number of tests required by reader $R$
to find a suitable random seed (in OPT-C1) with $96$-bit ID from the universe $U$,
in order to verify two facts: (1) the probability of finding a suitable one within $6$ tests is close to $1$;
and (2) the average number of tests required is under $2.5$
(i.e., to verify Theorem \ref{number_of_trials_4_find_random_seeds}).
In particular,
we compute how many random seeds should be tested in order to solve the basic sampling problem
of selecting at least $c$ and at most $c+6\sqrt{c}$ tags randomly from a single category $P$ of $n$ tags.
Any $n$-subset of $U$ can be $P$.
Given fixed values of $c$ and $n$, we conduct $10^3$ independent trials
to calculate the actual numbers of tested random seeds for
finding a suitable one.
Each trial includes $6$ steps:
\begin{enumerate}
\item[(1)] Initialize a variable $x$ to be $0$.
\item[(2)] Generate the category $P$ by randomly choosing  $n$
tags from the universe $U$ (i.e., choosing $n$ IDs from the $2^{96}$ different IDs).
\item[(3)] Pick a random seed $r$,
and compute the corresponding hash values $H(t,r)\bmod n = h(t)$ for each tag $t\in P$.\footnote{\textcolor{black}{Note: a random seed $r$ is actually the starting position of a $\log_2{(n)}$-bit substring in the $96$-bit ID of tag $t$, and
the decimal value of this substring is the hash value for $t$, see Theorem \ref{communication_cost_OPT-C1}.}}
\item[(4)] A tag $t\in P$ enters the selected state if $h(t)\le c+3\sqrt{c}$.
\item[(5)] Calculate the value of $l$, which represents the number of tags that enter the selected state
\item[(6)] If $l\in[c,c+6\sqrt{c}]$, stop the trial and report the value of $x$;
otherwise, set $x=x+1$ and return to step (3).
\end{enumerate}



We conduct $10^3$ trials in two typical scenarios, referred to as Scenario 7 and Scenario 8. These scenarios feature varying relationships between the parameters $n$ and $c$.
Scenario 7 maintains $n$ at $10^3$ and increases $c$ from $1$ to $10^2$,
while Scenario 8 keeps $c$ at $50$ and varies $n$ from $10^2$ to $10^3$.
The results for Scenarios 7 and 8 are presented in Fig. \ref{fig_for_sc5_experiment} and \ref{fig_for_sc6_experiment}, respectively.
\textcolor{black}{These figures reveal that the average number of tested random seeds is below $1.15$, and the maximum is always under $6$, thereby
verifying Theorem \ref{number_of_trials_4_find_random_seeds} in practice.
}

}}

{\color{black}{
\section{Experimental Results}\label{experimental_results}

In this section, we
present the results of our experiments aimed at
evaluating the real-world performance of the $\text{OPT-C}_\mathbf{IMPL}$
protocol against state-of-the-art
C1G2-compatible sampling protocols
within a real RFID system. These experiments are conducted using an Impinj R420 reader, denoted by  $R$, and a collection of $200$ COTS tags, specifically ALN-9630, NXP-Ucode7, and Impinj-ER62, all of which are compliant with the C1G2 standard \cite{EPCGlobal}.

Our experimental setup involves integrating the Impinj R420 reader with a 900 MHz Laird polarized antenna. This reader, $R$, operates in dense Miller 8 mode and is connected to a laptop computer equipped with an Intel i7-7700 processor and 16GB of RAM via an Ethernet connection. Varying quantities of tags, ranging from 40 to 200, are placed at a distance of approximately 1.5 meters from the antenna. For a visual representation of the physical configuration of the RFID system, please refer to Fig. \ref{fig_real_scenario_review_appendix}.
The implementation of $\text{OPT-C}_\mathbf{IMPL}$ is carried out in C\# using the Octane SDK .NET 4.0.0 \cite{ImpSDK}, running on the laptop to control the reader for tag recognition.

Our investigation of existing literature on RFID tag sampling \cite{8718354,hu2019bringing,yang2017analog,liu2018efficient,chu2021efficient} leads us to
identify the DTS protocol proposed in \cite{8718354} as the sole state-of-the-art tag sampling protocol that
can be directly implemented on COTS RFID devices
(i.e., fully C1G2-compatible).
The DTS protocol facilitates the random selection of tags from a subset with user-specified probabilities.
However, the DTS protocol cannot guarantee
to select precisely $c_k$ tags from each category $P_k$ ($k=1,2,\cdots,K$),
as each tag is independently chosen with a fixed probability. This leads to the possibility of selecting more than $c_k$ tags in practice (as discussed in section 4.2.2 of \cite{8718354}).
To address this, we enhance the DTS protocol by introducing a trimming mechanism to deselect surplus tags when more than $c_k$ tags are chosen from category $P_k$,
ensuring equitable comparisons.\footnote{\textcolor{black}{In the DTS protocol, reader $R$ begins by pre-writing a random bit array into the User memory of each tag in category $P_k$, ($k=1\sim K$). $R$ then employs a random selection method to choose tags from $P_k$. This selection process involves randomly picking a $\mlq1\mrq$ from the pre-written array, serving as the mask for the 'Select' command. By using this mask, $R$ can select multiple tags from category $P_k$.
To ensure the selection of exactly $c_k$ tags from
$P_k$, $R$ manages this by either deselecting surplus tags when more than $c_k$ tags are initially chosen from $P_k$ or selecting additional tags from $P_k$ through extra {\tt Select} commands when less than $c_k$ are initially chosen from $P_k$. This process allows reader $R$ to achieve the desired selection of $c_k$ tags from category $P_k$.
Once the selection of $c_k$ tags is successfully completed, reader $R$ proceeds to assign a unique reporting order to each selected tag. This assignment is facilitated by executing a standard inventory process, where each selected tag is associated with a singleton
slot (equivalently to assigning unique reporting orders).
}}
In addition to DTS, we design
a baseline tag sampling protocol called RandomSelect.
This protocol selects each sampled tag using an individual {\tt Select} command,
and is fully C1G2-compatible.\footnote{\textcolor{black}{The RandomSelect protocol begins with reader $R$ randomly selecting $c_k$ tags from category $P_k$
by drawing $c_k$ EPCs from the pool of EPCs associated with tags in $P_k$. Subsequently, reader $R$ executes $c_k$ individual {\tt Select} commands,
with each command utilizing the EPC of the corresponding selected tag as a mask.
Once the selection of $c_k$ tags is successfully completed, reader $R$ proceeds to assign a unique reporting order to each selected tag. This assignment is accomplished through the execution of a standard inventory process, where each selected tag is associated with a singleton slot.
}}

\begin{table}[!t]
\centering
{\color{black}{
    \caption{\color{black}Parameter settings for the
    three real scenarios}
    \label{tab:ps}
  \centering
  \begin{tabular}{p{1.2cm}!{\vrule width 1 pt}p{2.8cm}|p{5.4cm}|p{3.8cm}}
      \toprule[1.5pt]
       Real Scenario &The number of categories &Category sizes& Sample sizes  \\
      \hline
      \Rmnum{1} & $K$ $\in\{2, 4, ..., 10 \}$
      &
      $n_1=n_2=...=n_K=20$
      &
      $c_1=c_2=...=c_K=5$\\
      \hline
      \Rmnum{2} 
      & $K=8$ & $n_1=n_2=...=n_K\in\{10,15,20,25\}$ &
      $c_1=c_2=...=c_K=4$\\
      \hline
      \Rmnum{3} 
      &$K=8$&
      For each $k\in\{1,2,...,K\}$, if $k\bmod 2=1$, $n_k=20$;
        if $k\bmod 2 = 0$, $n_k=30$
      & $c_1=c_2=...=c_K\in\{2,4,6,8,10\}$
        \\
      \bottomrule[1.5pt]
    \end{tabular}
}}
\end{table}

\begin{figure*}[!t]
  \centering
  \subfigure[Real Scenario \Rmnum{1}]{
    \label{fig_for_sce1_time} 
    \includegraphics[width=0.24\linewidth]{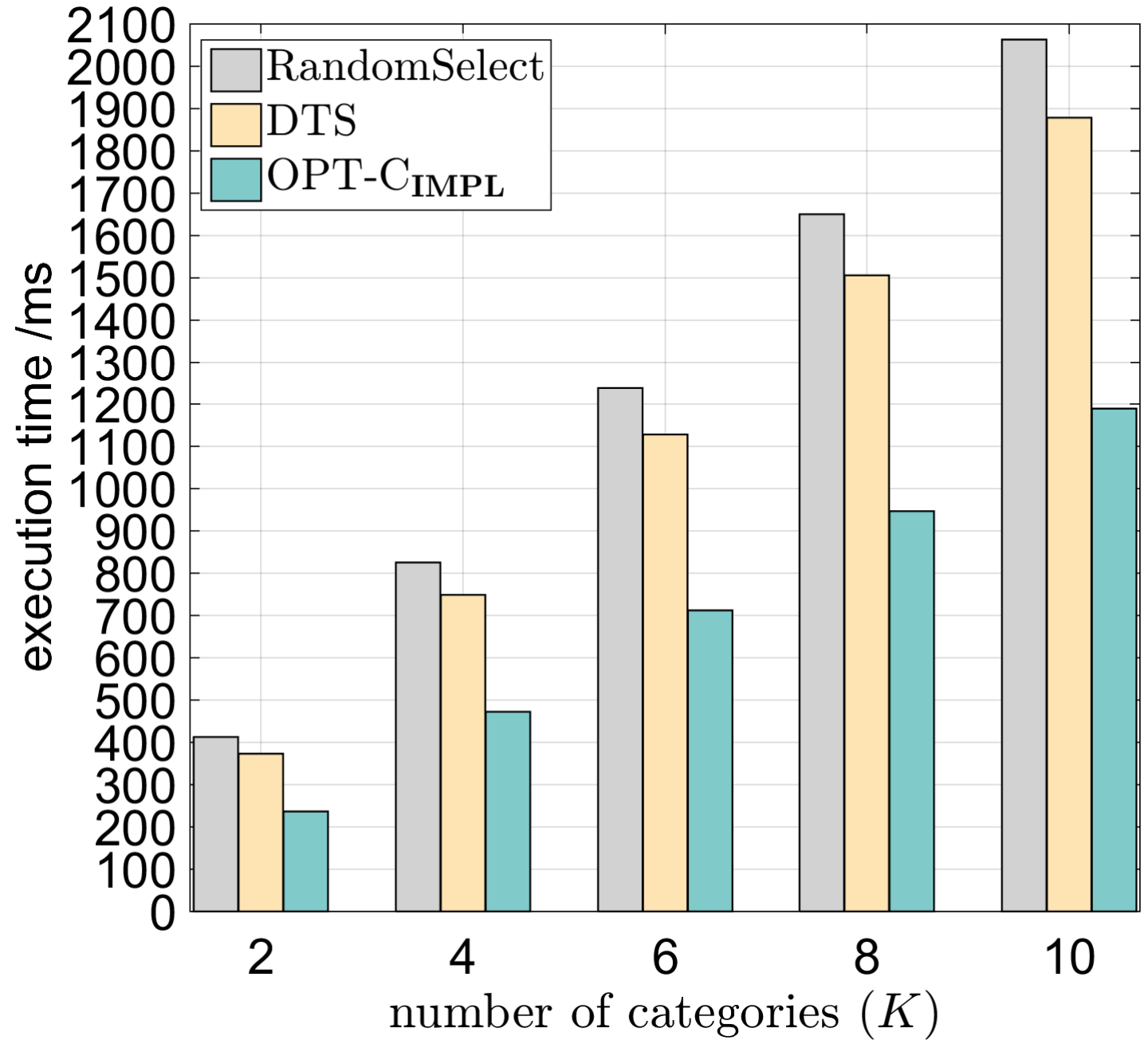}}
   \subfigure[Real Scenario \Rmnum{2}
   ]{
    \label{fig_for_sce2_time} 
    \includegraphics[width=0.24\linewidth]{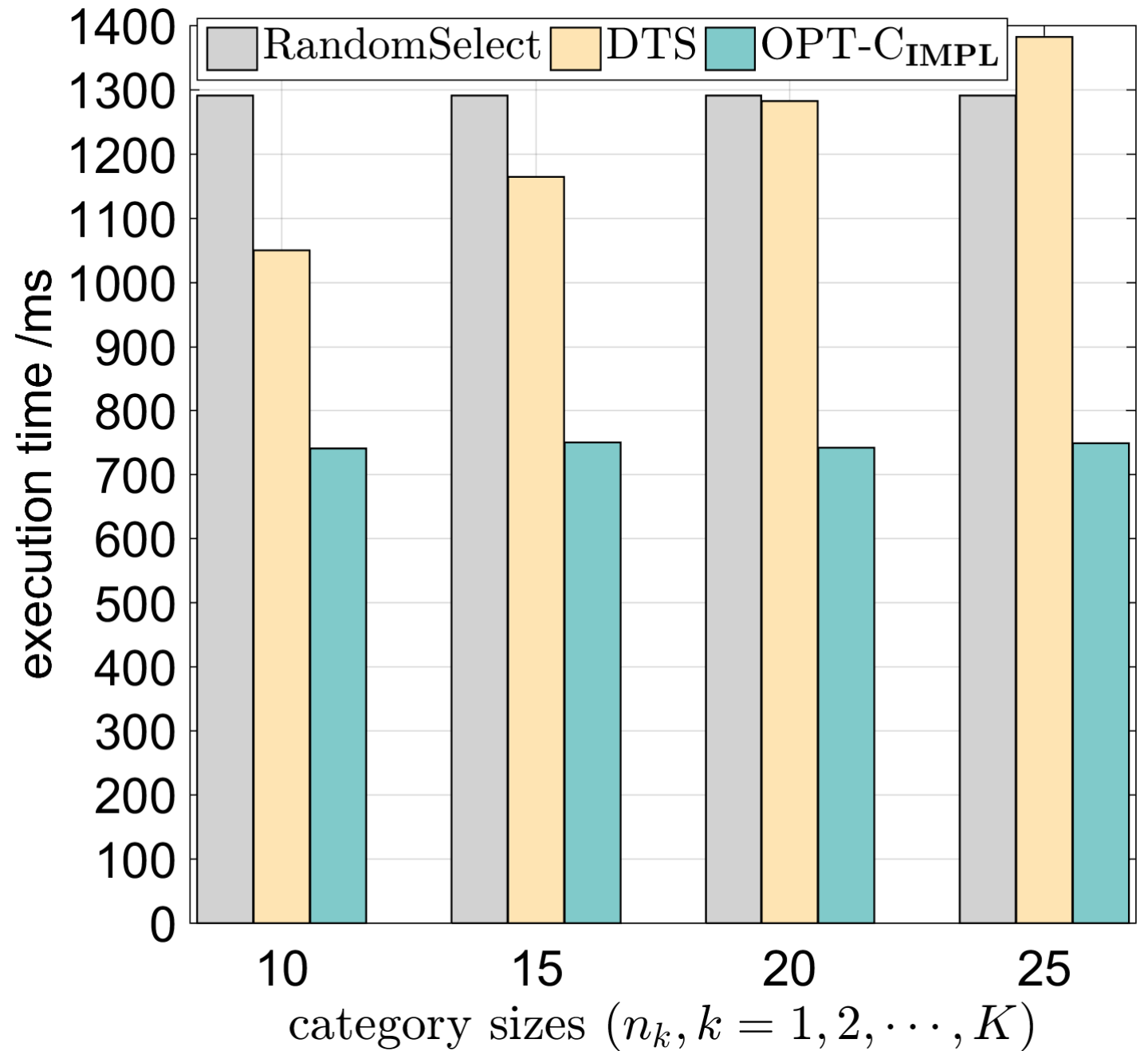}}
    \subfigure[Real Scenario \Rmnum{3}
    ]{
    \label{fig_for_sce6_time} 
    \includegraphics[width=0.24\linewidth]{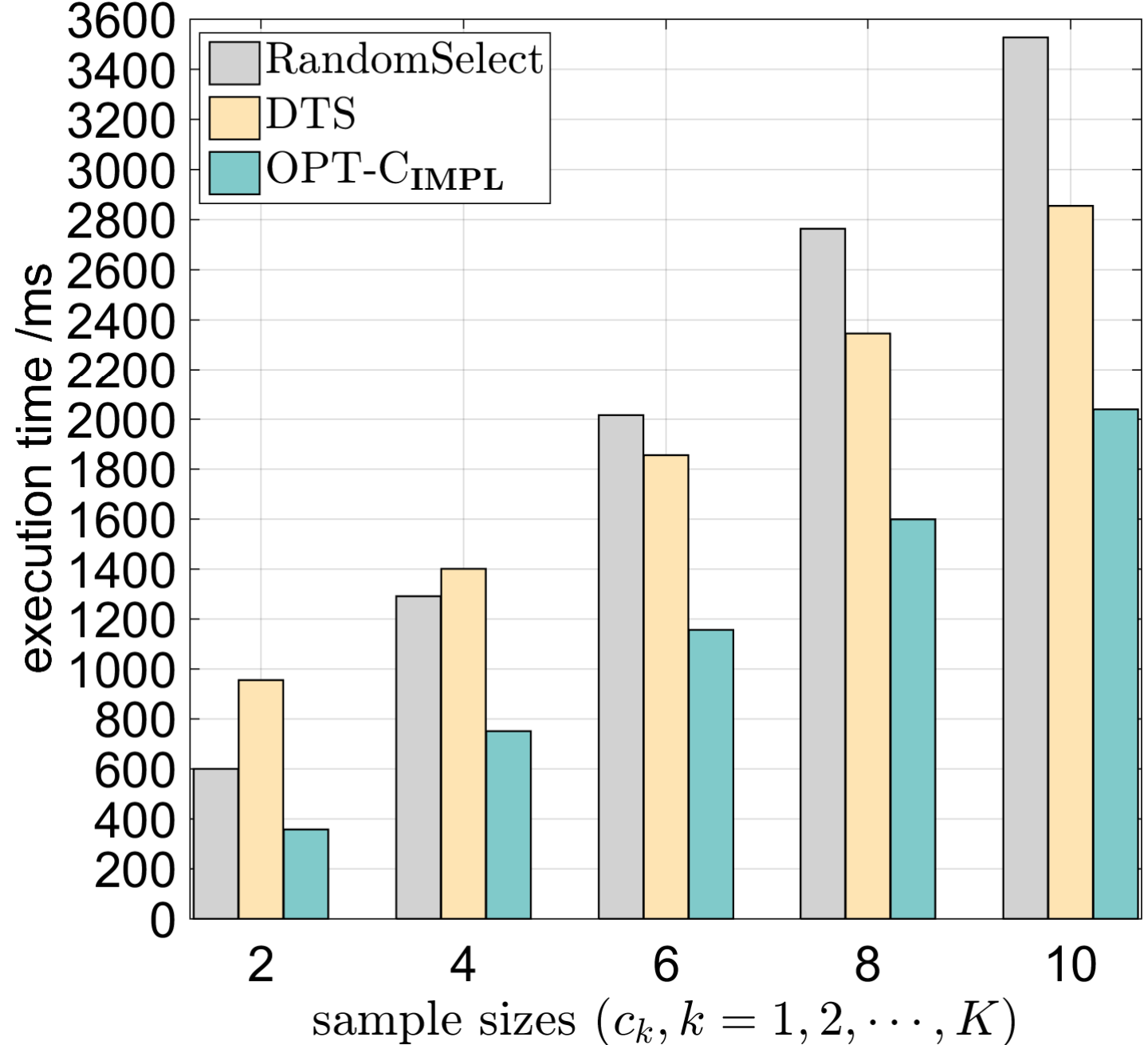}}
  \caption{\textcolor{black}{The practical communication time of RandomSelect, DTS,
  and $\text{OPT-C}_\mathbf{IMPL}$
  for Real Scenarios \Rmnum{1}, \Rmnum{2}, and \Rmnum{3}.}}
  \label{fig_fpr}
\end{figure*}

We access the performance of the three protocols (RandomSelect, DTS, and $\text{OPT-C}_\mathbf{IMPL}$)
in $3$ typical real scenarios
(Scenarios \Rmnum{1}-\Rmnum{3}).
Each scenario is characterized by variations in one of three key parameters, which include $K$ (the number of categories), $n_1, n_2, \cdots, n_K$ (category sizes), and $c_1, c_2, \cdots, c_K$ (sample sizes or reliability numbers).
The detailed parameter settings for Scenarios \Rmnum{1}-\Rmnum{3} are listed in Tab. \ref{tab:ps}.
For each scenario, we conduct 10 trials and record the average practical execution times of the three protocols.
Here's how we conduct the trials in each scenario:

\noindent$\bullet$ \emph{\textbf{Step 1:}  We manually pick $N=\sum\nolimits_{k=1}^K n_k$ tags from the universe $\mathcal{T}$ of $500$ COTS tags 
and divide them into $K$ categories: $P_1, P_2,\cdots, P_K$,
where $P_k$ contains $n_k$ tags ($k=1,2,\cdots,K$)
assigned with the same category-ID $k$.}

\noindent$\bullet$ \emph{\textbf{Step 2:}  We let the reader $R$ execute one of the three tested protocols over this tag population to sample $c_k$ tags from each category $P_k$ and assign unique integers to the sampled tags. We record the time taken by each protocol.}

\noindent$\bullet$ \emph{\textbf{Step 3:} The reader $R$ pauses and waits for us to manually select another $N$ tags from $\mathcal{T}$ for the next trial.}

The experimental results from these three real scenarios offer substantial evidence supporting the superior performance of $\text{OPT-C}_\mathbf{IMPL}$.
In particular, Fig. \ref{fig_fpr} demonstrates
that $\text{OPT-C}_\mathbf{IMPL}$ consistently outperformed DTS and RandomSelect in terms of average communication time.
For instance, in Scenario \Rmnum{1}, $\text{OPT-C}_\mathbf{IMPL}$ achieves approximately $63\%$ of
DTS's average communication time and $57\%$ of RandomSelect's average communication time.
Similar trends can be observed in Scenarios \Rmnum{3} and \Rmnum{5}, where $\text{OPT-C}_\mathbf{IMPL}$
outperforms DTS and RandomSelect, achieving around $62\%$ and $57\%$ of their respective average communication times.

The superior performance of $\text{OPT-C}_\mathbf{IMPL}$ can be attributed to the OPT-C protocol, which offers two key advantages:
\textbf{(1) Coarse Sampling Stage:} $\text{OPT-C}_\mathbf{IMPL}$ follows the first stage of OPT-C, employing a specially designed threshold and tag hash values. This allows $\text{OPT-C}_\mathbf{IMPL}$ to 
sample slightly more than $c_k$ tags from category $P_k$ using fewer than $\lceil\log_2(c_k)\rceil+1$ {\tt Select} commands.
\textbf{(2) Refined Sampling Stage:} $\text{OPT-C}_\mathbf{IMPL}$ follows the second stage of OPT-C, using a specifically designed hash value range and threshold. This enables $\text{OPT-C}_\mathbf{IMPL}$ to draw exactly $c_k$ tags from category $P_k$ and assign them unique reporting orders using fewer than $2\lceil\log_2(c_k)\rceil+2$ {\tt Select} commands.

In summary, the number of {\tt Select} commands used by $\text{OPT-C}_\mathbf{IMPL}$
for
\textcolor{black}{the CIS problem}
is at most $\sum\nolimits_{k=1}^K (3\lceil\log_2(c_k)\rceil+3)$,
with each {\tt Select} command employing a filter string of $O(\log_2(c_k))$ bits.
In contrast, the state-of-the-art DTS protocol \cite{8718354} demands approximately
$\sum\nolimits_{k=1}^K O(\sqrt{c_k})$ {\tt Select} commands
and an additional operation involving pre-writing a lengthy
bit-vector into each tag's user memory before sampling tags.\footnote{\textcolor{black}{By section 4.2.2 of \cite{8718354},
the DTS protocol samples each in $P_k$ independently with a fixed probability of
$\mathsf{p}_k=c_k / n_k$. Let $\mathsf{X}_k$ denote the number of sampled tags from $P_k$ by the DTS protocol.
Then $|\mathsf{X}_k-c_k|$ represents the additional tags that DTS protocol must further
select or unselect to ensure exactly $c_k$ are drawn from $P_k$.
By the second formula in Theorem 1 of \cite{berend2013sharp},
we know that $E[|\mathsf{X}_k-c_k|] \ge \sqrt{n_k \mathsf{p}_k
(1-\mathsf{p}_k)}/\sqrt{2} = \sqrt{n_k(c_k/n_k)(1-c_k / n_k)}/\sqrt{2}=O(\sqrt{c_k})$.
Hence, on average, DTS needs to execute $O(\sqrt{c_k})$ {\tt Select} command to
ensure exactly $c_k$ tags are drawn from category $P_k$.
}}
The RandomSelect protocol necessitates $\sum\nolimits_{k=1}^K c_k$ {\tt Select} commands, each utilizing a $96$-bit array as its mask.
}}

\section{Conclusions}\label{conclusion}
In this paper, we have studied the Category Information Sampling Problem
in practical RFID systems with missing tags.
Our primary objective is to use the shortest possible time
to randomly select a few tags from each group and collect their category information.
For this target,
we first obtained a lower bound on
the execution time for solving this problem. Then, we presented
a near-optimal protocol OPT-C, and proved that OPT-C
can solve the studied problem with an execution time
close to the lower bound.
Lastly, we verified the practicability and validity of OPT-C
with real experiments, and compared OPT-C with the
state-of-the-art solutions through extensive simulations
to demonstrate its superiority.
\appendix
\subsection*{A.$\quad$An analytical calculation of $c_k$ to
guarantee a given success probability}\label{appendix_reqc}
Assume that $\alpha_k\in(0,1)$ represents the rate of missing tags
in category $P_k$,
which can be estimated very rapidly with a time cost of only $O(\log_2(\log_2(|P_k|))$ \cite{zhou2014understanding}.
{When only one tag is selected randomly from $P_k$, the success probability is $1-\alpha_k$,
i.e., the probability that the randomly selected tag is not missing is $1-\alpha_k$.}
However, if we choose $c_k$ tags randomly from $P_k$, the success probability increases rapidly to
$1-\frac{(\alpha_k |P_k|)^{\underline{c_k}}}{|P_k|^{\underline{c_k}}}\ge 1-{\alpha_k}^{c_k}$.
This can be explained as follows:
as long as one of the $c_k$ randomly selected tags from $P_k$
is not missing, we can successfully collect $P_k$'s category information.
So the success probability is equal to the probability that at least one of the $c_k$
randomly selected tags is not missing (i.e., equal to $1-\frac{(\alpha_k |P_k|)^{\underline{c_k}}}{|P_k|^{\underline{c_k}}}$).
Furthermore,
given a small $\varepsilon\in(0,1)$,
to guarantee a high success probability of $1-\varepsilon$,
we only need to randomly choose $c_k=\ln(\frac{1}{\varepsilon})/\ln(\frac{1}{\alpha_k})$,
because ${\alpha_k}^{c_k}={\alpha_k}^{\ln(\frac{1}{\varepsilon})/\ln(\frac{1}{\alpha_k})}=\varepsilon$.
\textcolor{black}{In addition, based on the inequality (4) of Ref. \cite{topsoe2007some}:
$\ln(1+z) \le \frac{2z}{2+z}, \text{ $-1 < z \le 0$} $,
we can obtain that $\ln(\frac{1}{\alpha_k}) \ge \frac{2(1-\alpha_k)}{1+\alpha_k}$.
Subsequently,
since $c_k=\ln(\frac{1}{\varepsilon})/\ln(\frac{1}{\alpha_k})\le \ln(\frac{1}{\varepsilon})\frac{1+\alpha_k}{2(1-\alpha_k)}$,
we can observe that $c_k$ remains small even when $\alpha_k$ is relatively large and $\varepsilon$ is relatively small.}

\subsection*{B.$\quad$Proof of the inequality: $\eqref{number_of_partitions_handled_by_v1} \le \eqref{number_of_partitions_handled_by_v2}$}\label{appendix_4_maximal_value_analysis}
Clearly, \eqref{number_of_partitions_handled_by_v1} can be represented by
a function $\mathrm{g}(s^v_{1,0},s^v_{1,1},...,s^v_{k,c_k})$
of variables: $s^v_{k,i}$, $k\in\{1,2,\cdots,K\}$, $i\in\{0,1,\cdots,c_k\}$.
Note that category sizes $n_1,n_2,\cdots,n_K$ and reliability numbers $c_1, c_2,\cdots, c_K$
are pre-given.
Furthermore, $\mathrm{g}(s^v_{1,0},s^v_{1,1},...,s^v_{k,c_k})$
can be expressed as the product of $K$ functions: $f(s^v_{k,0},s^v_{k,1},\cdots,s^v_{k,c_k})=\frac{(s^v_{k,0})^{n_k-c_k}}{(n_k-c_k)!}\prod\nolimits_{i=1}^{c_k}s^v_{k,i}$, $k\in\{1,2,...,K\}$.
We can find the maximum value of each of the $K$ functions by
studying the following maximization problem.
\begin{align}\nonumber
&\max\,\, \textcolor{black}{\ln(f(s^v_{k,0},s^v_{k,1},\cdots,s^v_{k,c_k}))}\\
&s.t.\quad
\begin{cases}
\sum\nolimits_{i=0}^{c_k} s^v_{k,i}=\Upsilon-\sum\nolimits_{j=1}^{k-1}n_j, \nonumber
\\s^v_{k,i} \geq 0 \quad (i=0,1,\cdots,c_k).
\end{cases}
\end{align}
{\color{black}{
Clearly, the function $\ln(f(s^v_{k,0},s^v_{k,1},\cdots,s^v_{k,c_k}))=
\ln(\frac{(s^v_{k,0})^{n_k-c_k}}{(n_k-c_k)!})+\sum\nolimits_{i=1}^{c_k}\ln(s^v_{k,i})=
(n_k-c_k)\ln({s^v_{k,0}}) + \sum\nolimits_{i=1}^{c_k}\ln(s^v_{k,i})-\ln({(n_k-c_k)!})
$ is a non-negative linear combination of $c_k + 1$ concave functions:
$\ln({s^v_{k,0}})$, $\ln(s^v_{k,1})$, $\ln(s^v_{k,2})$, $\cdots$, $\ln(s^v_{k,c_k})$.\footnote{
Function $\ln(x)$ is concave since its second-order derivative is $\frac{{{d^2}\ln (x)}}{{d{x^2}}} = ( - 1)\frac{1}{{{x^2}}}$ which is below $0$.
}
Thus, $\ln(f(s^v_{k,0},s^v_{k,1},\cdots,s^v_{k,c_k}))$ is also concave \cite{cambini2008generalized}.
So we can use the method of Lagrange multipliers
to discover that its maximum value is
$({n_k-c_k})\ln((n-c_k)\frac{\Upsilon-\sum\nolimits_{j=1}^{k-1}n_j}{n_k})+
\sum\nolimits_{i=1}^{c_k}\ln(\frac{\Upsilon-\sum\nolimits_{j=1}^{k-1}n_j}{n_k})--\ln({(n_k-c_k)!})$,
which is
obtained when $s^v_{k,i}=\frac{\Upsilon-\sum\nolimits_{j=1}^{k-1}n_j}{n_k}$, $i=1,2,\cdots,n_k$,
and $s^v_{k,0}=(n-c_k)\frac{\Upsilon-\sum\nolimits_{j=1}^{k-1}n_j}{n_k}$.
Since function $\ln(x)$ is monastically increasing,
we know $$f(s^v_{k,0},s^v_{k,1},\cdots,s^v_{k,c_k})\le \frac{(n_k-c_k)^{n_k-c_k}}{(n_k-c_k)!}\big( \frac{\Upsilon-\sum\nolimits_{j=1}^{k-1}n_j}{n_k}\big)^{n_k}$$.
}}
Multiplying the maximum values of the $K$ functions together
gives us
$$\prod\nolimits_{k=1}^{K} \left[\frac{(n_k-c_k)^{n_k-c_k}}{(n_k-c_k)!}\left( \frac{\Upsilon-\sum\nolimits_{j=1}^{k-1}n_j}{n_k}\right)^{n_k}\right]=\Phi,$$
which is
the maximum of $\mathrm{g}(s^v_{1,0},s^v_{1,1},...,s^v_{k,c_k})$.
This leads to the inequality $\eqref{number_of_partitions_handled_by_v1} \le \eqref{number_of_partitions_handled_by_v2}$.

\subsection*{C.$\quad$Proof of the fact: $\mu(c_k)=e^{3\sqrt{c_k}}(1+\frac{3\sqrt{c_k}}{c_k+3\sqrt{c_k}})^{-c_k-6\sqrt{c_k}}$
is a 
monotonically decreasing function of $c_k$}\label{appendix_a}
Let $x=\sqrt{c_k}$,
then the function $\mu(c_k)$ 
can be rewritten as:
$\mu(x^2) =  e^{3x} (1+\frac{3x}{x^2+3x})^{-x^2-6x} =e^{3x + ({-x^2-6x})\ln(1+\frac{3x}{x^2+3x})}$.
Since $c_k \ge 1$ and $x\ge1$,
proving that the derivative of $\mu(x)$ is less than $0$ will
be enough to
show $\mu(c_k)$ is a monotonically decreasing function.
The derivative of $\mu(x)$ is analyzed as below:
\begin{flalign*}
&{d\,\mu(x)}/{d\,x} = 3 - (2x+6) \ln(1+{3x}/({x^2+3x})) \nonumber\\
&\, - ({x^2+6x})\frac{{x^2+3x}}{{x^2+6x}}\frac{(2x+6)(x^2+3x) - (x^2+6x)(2x+3) }{(x^2+3x)^2}\nonumber\\
&\le \frac{6x+9}{x + 3} -  \frac{12}{2+3/({x+3})} =\frac{ {{-27}} }{(x+3)(2x+9)} < 0. \nonumber
\end{flalign*}
The first inequality in the above derivation
is based on the inequality (3) of Ref. \cite{topsoe2007some}:
$  \ln(1+z) \ge \frac{2z}{2+z}, \text{$ 0 \le z < \infty $ }$.


\subsection*{D.$\quad$Proof of the fact:
$\rho(c_k)=e^{-3\sqrt{c_k}}(1+\frac{-3\sqrt{c_k}}{c_k+3\sqrt{c_k}})^{\textcolor{black}{-c_k}}$
is a monotonically decreasing function of $c_k$
}\label{appendix_b}
Let $x=\sqrt{c_k}$,
then the function $\rho(c_k)$ 
 can be rewritten as
$\rho(x^2)=e^{-3x + x^2(\ln(\frac{x^2+3x}{x^2}))}$.
So, since $c_k\ge 1$ and $x\ge 1$,
proving that the derivative of $\rho(x)$ is less than $0$ will be enough to
show $\rho(c_k)$ is a monotonically decreasing function.
The derivative of $\rho(x)$ is analyzed as below
\begin{flalign}
&{d\,\rho(x)}/{d\,x} = -3 + 2x\ln\big(\frac{x^2+3x}{x^2}\big)\nonumber\\
&\quad\quad\quad\quad\quad + x^2\frac{x^2}{x^2+3x}\frac{(2x+3)x^2 - 2x(x^2+3x)}{x^4}\nonumber\\
&\le -3 + 2x\left(\frac{\frac{3}{x}}{2}\cdot\frac{2+\frac{3}{x}}{1+\frac{3}{x}}\right) - \frac{3x}{x+3} =0.
\end{flalign}
The inequality in the above derviation
is from
the inequality (3) of Ref. \cite{topsoe2007some}:
$ \ln(1+z) \le \frac{z}{2} \cdot \frac{2+z}{1+z},\text{ $ 0 \le z < \infty $ }$.

{\color{black}{
\subsection*{E. The {\tt Select} command} 
According to the C1G2 standard, the reader $R$ is capable of using a {\tt Select}
command to pick a specific group of tags for the upcoming tag inventory.
Every tag has a flag variable {\tt SL}, and the {\tt Select} command
will turn the {\tt SL} of matched tags to {\tt asserted} (true) and
the {\tt SL} of unmatched tags to {\tt deasserted} (false).
Only the asserted tags will remain active and respond to reader $R$
in the upcoming tag inventory. A {\tt Select} consists of
six mandatory fields, which are explained below.\\
$\bullet$ The \underline{Target} field allows the reader to modify the tag's {\tt SL} flag or inventory flag.
    In $\text{OPT-C}_\mathbf{IMPL}$, the reader $R$ sets \underline{Target}$=100_2$ to change the {\tt SL} flag.\\
$\bullet$ The \underline{Action} field specifies the action to be executed by the tag. There are eight different action codes available in Table 6-30 of \cite{EPCGlobal}.
In $\text{OPT-L}_\mathbf{IMPL}$, the reader $R$ sets \underline{Action} to be $0$. This
means that the matched tags shall assert {\tt SL}, while the unmatched tags shall deassert {\tt SL}.\\
$\bullet$ The \underline{MemBank} field indicates the memory bank to which a tag will look for comparison with the \underline{Mask}. As the hash values are stored in \emph{EPC memory}, reader $R$ sets \underline{MemBank}$=1$.\\
$\bullet$ The \underline{Pointer} field records the starting bit in the selected \underline{MemBank}, and the \underline{Mask} will be compared to the bit string starting at this position.\\
$\bullet$
The \underline{Length} field specifies the length of the \underline{Mask} in bits.\\
$\bullet$ The \underline{Mask} is a user-specific bit-string that must match the content of a specific position in a \underline{MemBank}.

Overall, a tag will compare \underline{Mask} with the \underline{Length}-bit long string that starts at the \underline{Pointer}-th bit in its \underline{MemBank} to determine whether it is a matched or unmatched tag.

}}

{\color{black}{
\subsection*{F. Explanation of Lower Bound Factors}
The lower bound on execution time in Theorem 2.2 is predominantly
influenced by two key factors: the number of categories, denoted as $K$,
and the reliability numbers $c_1, c_2, \ldots, c_K$. Here, $K$ signifies
the number of different categories (disjoint subsets) in the tag population
$S$ and is a crucial parameter characterizing the distributional information of $S$.
 On the other hand, $c_k$ ($k\in\{1,2,\ldots,K\}$) specifies the number of tags that
 the system user intends to randomly draw from category $P_k$ and
assign unique reporting orders to. To elaborate on this result, consider
the following points:
\begin{itemize}
\item The reliability numbers $c_k$ ($k\in\{1,2,\ldots\}$) directly
impact the lower bound. The larger the number of tags that reader $R$
needs to draw from $P_k$ and assign unique reporting orders to, the
greater the amount of information $R$ must transmit to handle these selected tags.
For example, informing $100$ tags of a unique reporting order from $\{1,2,\ldots,100\}$
is typically more challenging than informing $10$ tags with unique
reporting orders from ${1,2,\ldots,10}$ since the former encompasses
the latter as a subproblem.
\item Another significant factor contributing to the lower bound is $K$,
representing the number of categories. As the reader needs to process an
increasing number of categories, more information must be transmitted by $R$
to handle each category effectively. For instance, sampling $1$ tag from each
of $100$ categories is generally more complex than sampling $1$
tag from each of $10$ categories, as the former includes the latter as
a subproblem.
\end{itemize}
Please note that the parameters $n_1, n_2, \ldots, n_K$ (representing the number of tags
contained in each of the $K$ categories) do not significantly influence the
lower bound. The fact that the lower bound on communication time mainly depends
on $K$, $c_1, c_2, \ldots, c_K$ indicates that the primary overhead of the sampling
problem potentially lies in the communication cost of assigning the $c_k$
randomly selected tags from $P_k$ unique reporting orders from $\{1,2,\ldots,c_k\}$
(as specified in Requirement (\textbf{\Rmnum{2}}) of Definition 2.1 in the manuscript).
This insight has further guided us to develop a two-stage protocol to address
the sampling problem in Section 3. In the first stage, roughly $c_k$ tags are
drawn from $P_k$ with minimal communication cost. Subsequently, in the second
stage, a unique reporting order is assigned to exactly $c_k$ of these previously
selected tags.
}}

\end{document}